\documentclass[sigplan,10pt]{acmart}

\usepackage[english]{babel}
\usepackage{blindtext}
\usepackage{amsmath} 
\DeclareMathOperator*{\argmin}{arg\,min}

\usepackage[ruled,vlined,linesnumbered]{algorithm2e}
\usepackage{wrapfig}
\usepackage[utf8]{inputenc}
\usepackage{booktabs}
\usepackage{caption}

\usepackage{booktabs}
\usepackage[table]{xcolor}
\usepackage{makecell}

\definecolor{YesBg}{RGB}{232,245,233}     
\definecolor{PartBg}{RGB}{255,248,225}    
\definecolor{NoBg}{RGB}{255,235,238}      

\definecolor{RowAlt}{RGB}{248,249,251}

\newcommand{\YES}{\cellcolor{YesBg}\textbf{Yes}}
\newcommand{\PART}{\cellcolor{PartBg}\textbf{Partial}}
\newcommand{\NO}{\cellcolor{NoBg}\textbf{No}}

\begin{document}

\copyrightyear{2026}
\acmYear{2026}
\setcopyright{cc}
\setcctype{by}
\acmConference[SIGOPS ATC 2026]{The 2026 ACM SIGOPS Annual Technical Conference}{November 15-18, 2026}{Hong Kong, China}
\acmBooktitle{The 2026 ACM SIGOPS Annual Technical Conference (SIGOPS ATC 2026), November 15-18, 2026, Hong Kong, China}
\acmDOI{10.1145/3828523.3857091}
\acmISBN{979-8-4007-2818-1/2026/11}

\title[Energy-Aware Scheduling for Serverless LLM Serving on Shared GPUs]{Energy-Aware Scheduling for Serverless LLM Serving on Shared GPUs (Extended)}

\makeatletter
 \let\originalACMReference\@mkbibcitation
 \renewcommand{\@mkbibcitation}{%
   \begingroup
   \let\@title\shorttitle
   \originalACMReference
   \endgroup
 }
 \makeatother
 
\author{Tianyu Wang}
\authornote{Part of the work was conducted during Tianyu's internship at HPE Labs.}
\affiliation{%
  \institution{University of Pittsburgh}
  \city{Pittsburgh} 
  \state{PA} 
  \country{USA}
}

\author{Gourav Rattihalli}
\affiliation{%
  \institution{HPE Labs}
  \city{Milpitas} 
  \state{CA} 
  \country{USA}
}

\author{Aditya Dhakal}
\affiliation{%
  \institution{HPE Labs}
  \city{Milpitas} 
  \state{CA} 
  \country{USA}
}

\author{Longfei Shangguan}
\affiliation{%
  \institution{University of Pittsburgh}
  \city{Pittsburgh} 
  \state{PA} 
  \country{USA}
}

\author{Dejan Milojicic}
\affiliation{%
  \institution{HPE Labs}
  \city{Milpitas} 
  \state{CA} 
  \country{USA}
}

\newcommand{\systemname}{\textsf{Festina}\xspace}
\newcommand{\systemnameposs}{\textsf{Festina'}\xspace}



\renewcommand{\shortauthors}{T. Wang et al.}
\begin{abstract}

As LLM inference becomes a major cloud workload, its growing energy footprint calls for cluster-wide optimization. Serverless serving handles traffic volatility by elastically sharing GPUs across models, but co-resident models share a device-wide operating point despite varying resource demands and latency slack across execution phases and loads. As a result, minimizing energy requires coordinated request placement, runtime resource adaptation, and workload consolidation.

We present \systemname, a profiling-guided, \emph{power-aware} control plane to minimize cluster-wide energy for serverless LLM serving. Unlike performance-oriented global--local schedulers, \systemname makes \emph{energy-first} decisions by jointly coordinating request placement, SM partitioning, and GPU operating points under TTFT/TBT SLOs. Its lightweight global scheduler enables fast, SLO-safe, energy-aware placement through constant-time lookups from offline profiles and GPU state summaries. On each GPU, a phase-aware local scheduler continuously adapts task batching and compute resources to minimize power.
\systemname further performs energy-aware workload consolidation to reduce GPUs' static power via SLO-aware migration.
Comparison with four SOTA LLM serving systems and one DVFS-augmented system demonstrates that \systemname reduces energy consumption by up to 56\% while maintaining parity in SLO attainment.

\end{abstract}



\maketitle

\definecolor{mygray}{gray}{0.9}
\newcommand{\boxC}[1]{%
  \begingroup
  \setlength{\fboxsep}{6pt}
  \noindent
  \colorbox{mygray}{\parbox{\dimexpr\linewidth-2\fboxsep\relax}{#1}}%
  \par\endgroup
}

\newcommand{\squishlist}{
    \begin{list}{$\bullet$}
        { \setlength{\itemsep}{0pt}      \setlength{\parsep}{0pt}
            \setlength{\topsep}{0.5pt}       \setlength{\partopsep}{0pt}
            \setlength{\listparindent}{-2pt}
            \setlength{\itemindent}{-5pt}
            \setlength{\leftmargin}{0.7em} \setlength{\labelwidth}{0em}
            \setlength{\labelsep}{0.2em} } }
    
\newcommand{\squishend}{
\end{list}  }

\definecolor{asparagus}{rgb}{0.55, 0.71, 0.0}

\newcounter{squishnum} 

\newcommand{\squishenum}{
    \begin{list}{\arabic{squishnum}.} 
        { \usecounter{squishnum}      
          \setlength{\itemsep}{0pt}
          \setlength{\parsep}{0pt}
          \setlength{\topsep}{0.5pt}
          \setlength{\partopsep}{0pt}
          \setlength{\listparindent}{-2pt}
          \setlength{\itemindent}{-5pt}
          \setlength{\leftmargin}{0.7em} 
          \setlength{\labelwidth}{0em}
          \setlength{\labelsep}{0.2em} } }

\newcommand{\squishenumend}{
\end{list}  }

\newcommand{\teal}[1]{\textcolor{teal}{#1}}
\newcommand{\blue}[1]{\textcolor{blue}{#1}}

\vspace{-5mm}
\section{Introduction}
\label{s:intro}

Large language model (LLM) inference is gradually becoming a dominant workload in today’s cloud~\cite{jaiswal2025serving}. Recent studies show that these inference workloads exhibit prominent spatial and temporal variation across regions, tenants, and time of day~\cite{muxserve,xia2025skylb,jaiswal2025serving,wang2025burstgpt,xu2025green}. To cope with this volatility, major platforms such as Microsoft Azure~\cite{msft_azure} and HuggingFace~\cite{hugging_face} now offer serverless LLM serving,  where the platform scales GPU resources up and down automatically to save inference cost without violating latency service level objectives (SLOs)~\cite{fu2024serverlessllm}. 
To further improve GPU utilization under fluctuating demand, there is a growing trend toward co-hosting multiple LLMs on the same GPU to multiplex requests with diverse workload characteristics~\cite{dhakal2020gslice,tan2021serving,muxserve,aegaeon}. 

At serverless scale, the energy footprint of LLM inference is projected to rise sharply as GPU fleets expand~\cite{AI_energy_consumption_MIT_review}. Prior study shows that for ChatGPT-like services, inference can dominate lifecycle emissions, producing around $25\times$ the carbon emissions of training a GPT-3 class model over a year~\cite{chien2023reducing}. A key driver is that modern accelerators draw high power when operated near their peak performance point. For instance, an NVIDIA H100 is configurable up to 700~W Thermal Design Power (TDP)~\cite{nvidia_hopper_architecture}. This makes power-efficient serving increasingly important: {\it small per-GPU inefficiencies quickly compound into a large cluster-wide energy bill.}

In this paper, we ask: \emph{how should a serverless LLM scheduler minimize cluster energy while preserving strict TTFT/TBT SLOs in a GPU-sharing pool?} Our starting point is an empirical opportunity: existing serving stacks often keep GPUs near their high-power operating region even when workload intensity is low or medium (\S\ref{sec:motivation}). As a result, many requests complete before Time To First Token (TTFT) and Time Between Tokens (TBT) deadlines, leaving substantial slack (\S\ref{sec:motivation_subsection_energy_opportunity}). 

This slack is not merely an artifact of our testbed. Production LLM workloads exhibit pronounced peaks and valleys due to bursty, long-tail demand~\cite{dynamollm,servegen,muxserve,li2024star}. For example, Azure Conversation traffic has a peak load 1.7$\times$ its average and 3.3$\times$ its valley, while Azure Coding reaches 2.8$\times$ and 34.6$\times$, respectively~\cite{dynamollm}. Because serving systems have to provision enough capacity to absorb peaks and preserve reliability, valley and medium-load periods naturally leave latency headroom for energy-aware scheduling.


At first glance, Dynamic Frequency and Voltage Scaling (DVFS) is the most immediate mechanism for exploiting this opportunity: when a GPU has latency headroom, the serving system can lower its frequency operating point and trade a portion of that headroom for lower power.
Nevertheless, the serverless, shared-GPU setting creates three forms of coupling that make simple per-instance DVFS insufficient.

\noindent $\bullet$ \textbf{Spatial coupling from co-location.}
In serverless, GPU-sharing setups, multiple LLMs are co-hosted on the same GPU and share one device-wide frequency operating point. 
However, different models and request intensities differ drastically in their individual energy-minimizing configurations (\S\ref{sec:motivation_u_shape}). As a result, a high-frequency-preferring workload will force a low-frequency-preferring workload to run at an unnecessarily high frequency, wasting significant energy (10\%-16\% energy waste in \S\ref{sec:motivation_u_shape}). Conversely, lowering the frequency to save energy for one model can endanger the deadlines of another co-resident model. Therefore, the scheduler should not focus on per-GPU or per-model frequency scheduling, but account for model placement, frequency affinity, SM availability, and queue state together.

\noindent $\bullet$ \textbf{Temporal coupling from prefill/decode dynamics.}
An LLM request shifts from compute-heavy prefill to memory-bound decode, and these two phases have different latency sensitivity, resource demand, and energy-optimal Streaming Multiprocessor (SM) and frequency configurations (\S\ref{sec:motivation_u_shape}). Moreover, the live batch on a GPU changes continuously as requests arrive, finish prefill, enter decode, and complete generation. A placement decision that is energy-efficient at dispatch time can quickly become suboptimal or even unsafe as the phase mix changes (\S\ref{sec:necessity_of_reconfiguration}). Energy-aware serving therefore requires runtime adaptation of task batching, SM partitioning, and the shared operating point, rather than a one-shot frequency decision.

\noindent $\bullet$ \textbf{Cluster-level coupling from serverless scale-in.}
During load valleys, consolidating work onto fewer GPUs can save substantial static and memory-system power. But consolidation is not ordinary bin packing. Moving a model would change the co-location set on both source and destination GPUs, which can recreate frequency mismatch and alter the energy-SLO tradeoff. If active requests are moved, transferring model state or KV cache also consumes time, network bandwidth, and energy. Thus, scale-in should be triggered only when the saved idle power outweighs migration overhead and when the resulting placement remains compatible in both resource demand and preferred operating point.

Therefore, the core challenge is not to
choose a frequency for an isolated model instance, but to coordinate placement,
SM partitioning, phase-aware runtime adaptation, and scale-in under a shared
GPU operating point.

\noindent We present \systemname, a two-tier control plane for energy-efficient serverless LLM serving on shared GPUs. At the cluster level (\S\ref{ss:design_freq_aware_dispatching}), a lightweight {\it global scheduler} performs queue-aware, SLO-safe request placement. Rather than simply selecting the least-loaded GPU, it chooses a feasible GPU whose current operating point, available SMs, memory capacity, and queued work best match the incoming request. This avoids placing requests on GPUs where they would either violate SLOs or create unnecessary operating-point mismatch.
However, as an LLM request quickly transitions from compute-heavy prefill to memory-bound decode, and the batch on a GPU changes as other requests arrive and finish, a request placement that is energy optimal at dispatch time can soon become suboptimal mid-execution, wasting energy and increasing SLO violation risk when local contention deviates from what the global scheduler assumed.

Festina bridges this gap with per-GPU {\it local scheduler} that continuously observes the live batch composition and jointly decides which prefill and decode tasks to run, how to partition SMs among them, and which shared operating point to use (\S\ref{ss:design_fine-grained_adaptation}).
This runtime adaptation lets \systemname track prefill/decode shifts and workload drift without relying on static per-model configurations. To keep control-plane overhead low at serverless scale, both global and local schedulers use an offline-profiled look-up table (LUT) that maps request features and resource configurations to latency and power (\S\ref{ss:design_offline_profiling}). At runtime, energy and latency reasoning reduces to constant-time table queries plus lightweight state checks.

To further minimize the substantial power consumption during periods of low GPU utilization, we propose an active scale-in mechanism where the global scheduler periodically evaluates cluster-wide load to identify opportunities for workload consolidation and adaptively triggers an LLM replacement to deactivate underutilized GPUs by co-locating models with similar optimal frequencies (\S\ref{ss:design_LLM_replacement}). 
To avoid disrupting ongoing inference, we further design a latency-aware migration algorithm that moves active requests only when their remaining execution time exceeds migration overhead.

\begin{table}[t]
\centering
\setlength{\tabcolsep}{5.5pt}
\caption{Comparison to related works.}
\vspace{-4.5mm}
\footnotesize

\rowcolors{2}{RowAlt}{white} 

\begin{tabular}{lcccc}
\toprule
& \makecell{\textbf{Spatial}\\\textbf{sharing}}
& \makecell{\textbf{Dynamic}\\\textbf{reconfiguration}}
& \makecell{\textbf{Energy}\\\textbf{aware}}
& \makecell{\textbf{Phase}\\\textbf{aware}} \\
\midrule
ServerlessLLM~\cite{fu2024serverlessllm} & \NO  & \PART & \NO  & \NO \\
MuxServe~\cite{muxserve}                 & \YES & \NO   & \NO  & \NO \\
Prism~\cite{yu2025prism}                 & \YES & \PART & \NO  & \NO \\
GreenLLM~\cite{liu2025greenllm}          & \NO  & \NO & \YES & \YES \\
DynamoLLM~\cite{dynamollm}               & \NO  & \PART & \YES & \NO \\
Dilu~\cite{dilu}                         & \YES & \YES  & \NO  & \NO \\
Aegaeon~\cite{aegaeon}                   & \YES & \YES  & \NO  & \PART \\
VoltanaLLM~\cite{yu2025voltanallm}       & \NO & \NO & \YES & \YES \\
\midrule
\systemname~(\textbf{ours})              & \YES & \YES  & \YES & \YES \\
\bottomrule
\end{tabular}
\label{tab:comparison}
\vspace{-6mm}
\end{table}

Note that \systemname~does not rely on a fixed resource configuration, such as the set of active GPUs, SM partitions, or operating frequencies.
Instead, it jointly determines request dispatching and runtime resource allocation to make use of most of the available slack while meeting SLOs.
As new requests enter the queue, the measured slack decreases.
In response, \systemname~allocates more resources and/or increases operating to accommodate the growing workload without violating SLOs.

We implement \systemname based on vLLM \cite{vLLM_2026}. 
Experiments based on industrial traces show that \systemname can save up to 56\% energy while maintaining parity in SLO attainment (within a 2\% margin) in comparison to the state-of-the-art (SOTA) LLM serving systems. 
To put this in perspective, in an industry AI cloud of 10,000 NVIDIA H100 GPUs, \systemname could potentially save over 2.8 million kWh of electricity per month: an amount equivalent to the monthly energy consumption of approximately 2,600 average U.S. households.
Sensitivity analysis confirms the robustness of \systemname across diverse workload compositions (e.g., prefill-heavy, decode-heavy) and validates its feasibility of energy-efficient execution under a prefill-decode disaggregated architecture.

To the best of our knowledge, \systemname is the first serverless LLM serving system that explicitly targets \emph{cluster-wide energy minimization} under a GPU-sharing regime by coordinating request placement, phase-aware task scheduling, SM partitioning, shared operating-point control, and scale-in migration under strict TTFT/TBT SLOs. This paper makes the following contributions:

\squishlist{}
    \item We formulate energy-efficient serverless LLM serving on shared GPUs as a coupled scheduling problem, where request placement, SM partitioning, phase-aware batching, shared operating-point control, and consolidation should be jointly decided under TTFT/TBT SLOs.
    \item We design \systemname, a multi-timescale control plane that combines queue-aware global placement, phase-aware local batching, joint SM/operating-point adaptation, and frequency-affinity-aware consolidation with migration-cost checks and efficient transitions.
    \item We implement \systemname on vLLM and evaluate it on an H100 testbed with industrial LLM serving traces. 
    \systemname cuts cluster energy consumption by up to 56\% over SOTA serving systems while preserving SLO attainment within 2\%.
\squishend{}

\vspace{-2mm}
\section{Motivation: Energy Coupling in Shared-GPU LLM Serving}
\label{sec:motivation}

In this section, we conduct measurement studies to explain why energy-efficient serverless LLM serving cannot be reduced to standalone DVFS. We first show that existing serverless GPU-sharing systems leave substantial SLO slack while GPUs remain near their high-power operating region, which creates an opportunity for energy-aware control.
We then demonstrate that GPU frequency affects energy in the expected U-shaped manner, consistent with prior DVFS studies~\cite{yu2025voltanallm,dynamollm,liu2025greenllm}.
However, our key finding is that the energy-optimal configuration shifts with the model, workload density, execution phase (i.e., prefill/decode), and SM allocation.
These shifts make energy optimization a coupled scheduling problem over model placement, SM partitioning, phase-aware runtime adaptation, and just-in-time scale-in.


\begin{figure}[t]
    \centering  
    \includegraphics[width=\columnwidth]{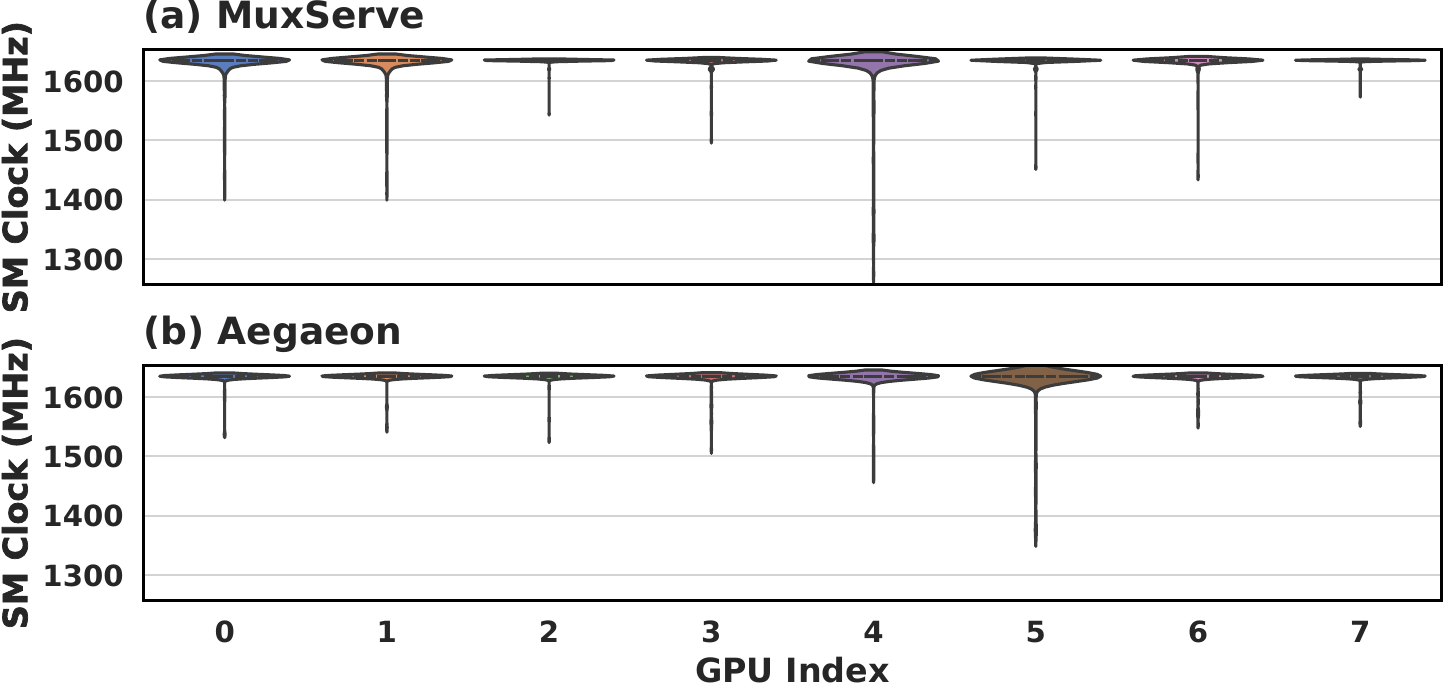}
    \vspace{-6mm}
    \caption{The frequency distribution of eight GPUs in 
    MuxServe~\cite{muxserve}  and Aegaeon~\cite{aegaeon} setups. \normalfont They all stay at a high 1635 MHz throughout the model serving session.
    }
    \label{fig:motivation_frequency_violin}
    \vspace{-2mm}
\end{figure}

\subsection{SLO Slack Creates Room for Energy-Aware Scheduling}
\label{sec:motivation_subsection_energy_opportunity}

We begin by examining how existing serverless GPU-sharing systems use GPU operating points under realistic serving workloads. We run MuxServe~\cite{muxserve} and Aegaeon~\cite{aegaeon} using MuxServe's traces and record GPU frequencies across eight NVIDIA H100 GPUs in a serverless serving setup. Figure~\ref{fig:motivation_frequency_violin} shows a clear pattern: under the default hardware policy used by these systems, GPUs frequently remain close to the TDP-limited frequency of 1635~MHz, despite substantial variation in the workload intensity.

\begin{figure}[t]
    \centering  
    \includegraphics[width=1\columnwidth]{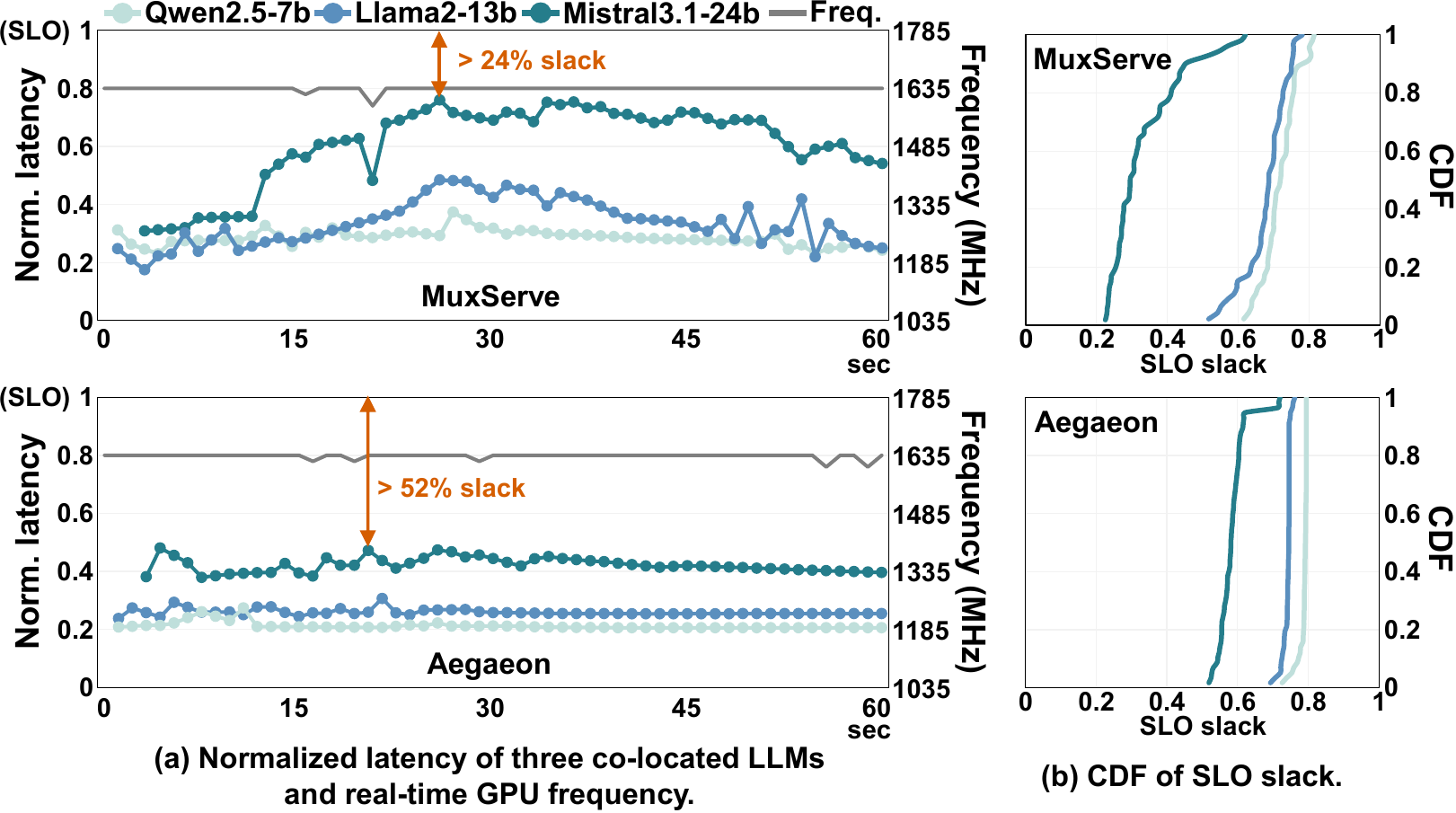}
    \vspace{-8mm}
    \caption{(a) Normalized latency of three co-located LLMs and real-time GPU frequency. (b) CDF of SLO slack for three LLMs during serving requests.}
    \label{fig:motivation_SLO_frequency}
    \vspace{-5mm}
\end{figure}

To understand whether this high operating point is necessary for meeting latency constraints, we measure request latency under the same setup. Figure~\ref{fig:motivation_SLO_frequency}(a) shows the normalized serving latency of three co-located LLMs, where a value below one means that the request completes within its SLO. Across the trace, all three models remain well below the SLO boundary. Under MuxServe, even the heaviest model, Mistral-24B, remains at least 24\% below the SLO limit at its worst point; Qwen-7B and Llama2-13B exhibit even larger margins for most of the trace. Aegaeon shows greater slack, with more than 52\% headroom, due to its prefill-decode disaggregated architecture. 

Figure~\ref{fig:motivation_SLO_frequency}(b) further shows that this slack persists throughout the trace rather than appearing only during short idle intervals. This behavior suggests an important opportunity. Existing serving stacks are optimized primarily for throughput, utilization, or SLO attainment, and typically leave GPU operating-point control to default hardware policies. As a result, the GPU can continue operating near a high-power point even when requests already finish well before their TTFT/TBT deadlines. Such unused latency headroom creates room for energy-aware scheduling: the serving system can trade part of the slack for lower power, as long as doing so does not increase queuing delay enough to violate SLOs.

\boxC{\textbf{Observation One:}
\emph{Serverless GPU-sharing workloads often have substantial SLO slack while GPUs remain near high-power operating points, creating room for energy-aware scheduling.}
}

\subsection{Energy-Optimal Operating Points Shift Across Workloads and Phases}
\label{sec:motivation_u_shape}

The previous result shows that serverless LLM requests often have latency slack. A natural next question is whether changing the GPU operating point can safely convert this slack into energy savings. Following common DVFS characterization methodology~\cite{dynamollm,xu2025green,yu2025voltanallm}, we profile phase-wise energy consumption of representative LLMs under different GPU frequencies and workload intensities. We use the same models as before and measure both prefill and decode energy under different input/output lengths.

\begin{figure}[tbp]
    \centering  
    \includegraphics[width=1\columnwidth]{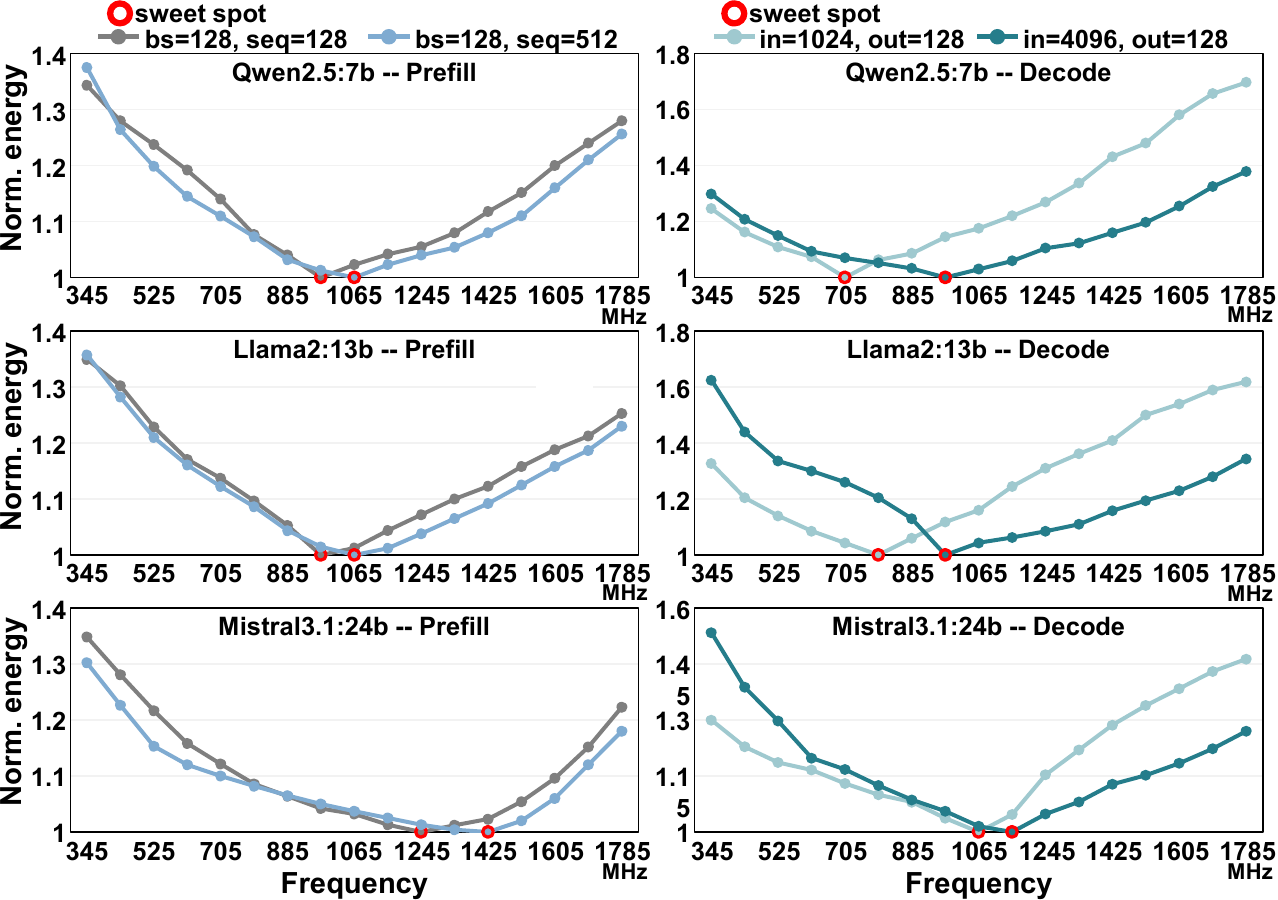}
    \vspace{-6mm}
    \caption{Impact of GPU frequency on energy consumption. 
    \normalfont
    Across all three LLMs, energy consumption follows a U-shaped curve during both prefill and decode. Red circles mark the energy-optimal frequency, while annotations indicate the remaining SLO slack at those operating points.}
    \label{fig:motivation_energu_frequency_pattern}
    \vspace{-8mm}
\end{figure}

Figure~\ref{fig:motivation_energu_frequency_pattern} shows that energy follows the expected U-shaped trend with respect to GPU frequency. At low frequencies, execution time increases and dominates total energy. While at high frequencies, power increases and dominates total energy. The minimum (i.e., sweet spot) appears at an intermediate frequency, reflecting the classic power--time tradeoff:
\begin{equation}\label{eq:energy}
    Energy = Power \times Time
\end{equation}

This U-shaped behavior is consistent with prior DVFS studies~\cite{dynamollm,xu2025green,yu2025voltanallm} and selecting the energy-minimizing operating point yields significant savings in Figure~\ref{fig:motivation_energu_frequency_pattern}: at least 16\% for prefill and 27\% for decode compared to the TDP limit, while still satisfying latency SLOs.

However, the real challenge is that for shared-GPU serving, the sweet spot is not fixed. We conclude three sources of variation from Figure~\ref{fig:motivation_energu_frequency_pattern}:

\squishlist{}
\item \textbf{Workload sensitivity.} For the same LLM, as workload intensity increases, e.g., with longer input sequences, the energy-minimizing operating point shifts toward higher frequencies because the workload becomes more compute-intensive and benefits more from higher throughput.

\item \textbf{Phase sensitivity.} The prefill phase generally prefers a higher frequency than the decode phase for the same LLM, which aligns with the compute-heavy nature of prefill and the memory-bound nature of decode~\cite{liu2025greenllm,wang2026prr}.

\item \textbf{Model sensitivity.} Larger models tend to favor higher frequencies than smaller models under comparable serving conditions because they impose heavier compute demand.
\squishend{}

Accordingly, when multiple LLMs are co-hosted on the same GPU, each individual model's energy sweet spot can conflict because the GPU exposes one device-wide operating point.
For instance, forcing Qwen2.5 and Llama2 to run at Mistral's optimal GPU clock increases their energy consumption by 10--16\%.
As a result, an operating point that saves energy for one model or phase may waste energy for another, while an overly aggressive downshift may slow co-resident work enough to threaten SLOs.

\boxC{\textbf{Observation Two:}
\emph{Consistent with prior DVFS studies, LLM serving exhibits a U-shaped energy-frequency tradeoff. The key challenge in serverless GPU sharing is that the energy sweet spot shifts across models, phases, and workload intensity, creating conflicts among co-resident workloads that share one device-wide operating point.}
}

\subsection{Frequency Scaling Alone Is Insufficient Under GPU Sharing}
\label{sec:necessity_of_reconfiguration}

The previous subsection shows that frequency is an important actuator. We now ask whether frequency control alone is sufficient once multiple LLMs share a GPU. In a GPU-sharing setting, SM partitioning is another primary determinant of execution time because it controls how much compute budget each co-resident workload receives. Frequency and SM allocation are therefore coupled: lowering frequency reduces dynamic power but can increase execution time, while changing SM allocation changes the latency impact of any frequency choice. To quantify this coupling, we compare three configurations that progressively expand the optimization scope: 
\squishlist{}
\item \textbf{S1: Performance-first baseline.} We select the SM partitioning that minimizes execution time and leave GPU frequency under default hardware control. 
\item \textbf{S2: Frequency-optimized.} We keep the latency-optimal SM partitioning from S1, but tune GPU frequency to minimize energy while preserving SLOs. 
\item \textbf{S3: Jointly optimized.} We jointly select SM partitioning and GPU frequency to minimize energy under the same SLO constraints. 
\squishend{}

\begin{figure}[t]
    \centering  
    \includegraphics[width=1\columnwidth]{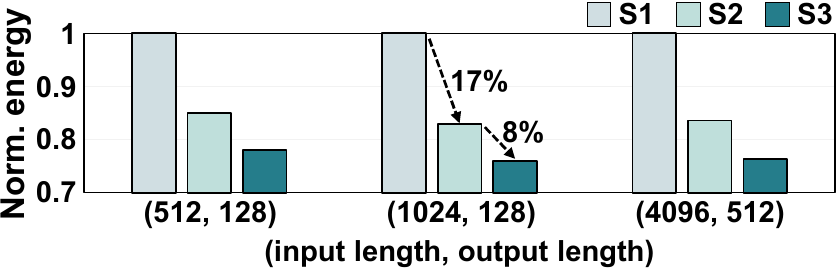}
    \vspace{-8mm}
    \caption{The normalized energy for three distinct settings under various workloads (i.e., input length).}
    \label{fig:motivation_GPU_sharing}
    \vspace{-6mm}
\end{figure}

Figure~\ref{fig:motivation_GPU_sharing} plots normalized energy under these configurations. Comparing S1 and S2 confirms that frequency scaling remains useful in GPU-sharing scenarios, reducing energy by 17\% in our measured setting. However, comparing S2 and S3 shows that frequency scaling alone leaves additional savings on the table: jointly reconfiguring SM partitioning and frequency reduces energy by another 8\%. This additional gain appears because changing SM allocation shifts the latency-energy tradeoff and can move the energy-minimizing frequency itself. 

More importantly, our experiments also show that the energy-optimal configuration is not static.
For instance, under the workload (1024 input tokens, 128 output tokens), the ideal (SM partitioning, frequency) setting shifts from (30-30-40 SMs, 1155 MHz) in S2 to (20-40-40 SMs, 1065 MHz) in S3. 
In real-world LLM serving, workloads are inherently dynamic: request rates fluctuate and token lengths vary continuously.

\boxC{\textbf{Observation Three:} Significant energy-saving potential exists in GPU-shared environments, but maximizing efficiency requires the co-reconfiguration of SM partitioning and frequency and the adaptation to workloads. 
}

\vspace{-2mm}
\section{\systemname Design}
\label{s:design}
\vspace{-1mm}



\begin{figure}[tbp]
    \centering  
    \includegraphics[width=1\linewidth]{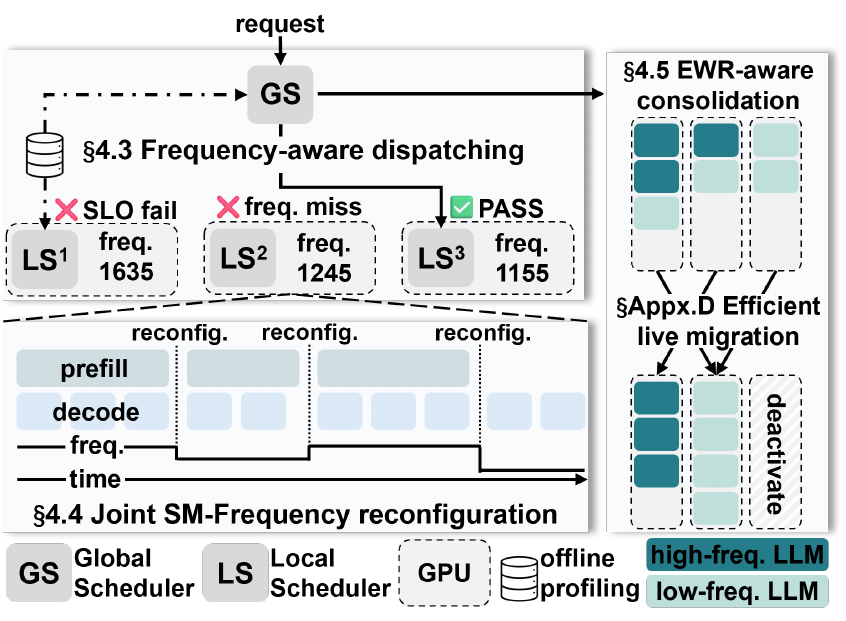}
    \vspace{-8mm}   
    \caption{System overview of \systemname. \systemname adopts a two-tier hierarchy for energy-aware serverless LLM serving.}
    \label{fig:overview}
    \vspace{-4mm}
\end{figure}

As shown in Figure~\ref{fig:overview}, \systemname uses a two-tier hierarchy for \emph{energy-aware} serverless LLM serving: a lightweight global scheduler makes fast, SLO-safe \emph{energy-aware} placement decisions across the GPU pool, while a per-GPU local scheduler performs phase-aware runtime adaptation by jointly tuning \emph{SM partitioning and GPU operating point} to save energy. Both tiers are unified by the same offline-profiled look-up table (\S\ref{ss:design_offline_profiling}), turning energy and latency reasoning into constant-time queries that scale with cluster size. 

\vspace{-1mm}
\subsection{System Overview}
\label{ss:design_overview}


\noindent$\bullet$ \textbf{Global Tier: Light-Weight, Frequency-Aware Dispatching (\S\ref{ss:design_freq_aware_dispatching}).} 
On each request arrival, the global scheduler uses the LUT and compact per-GPU state summaries to select an SLO-feasible GPU whose \emph{current} operating point best matches the request's predicted intensity. This keeps dispatch overhead low while avoiding energy waste from frequency-mismatched execution.

\noindent $\bullet$ \textbf{Local Tier: Joint SM-Frequency, Phase-Level Adaptation (\S\ref{ss:design_fine-grained_adaptation}).}
Because an LLM request shifts from compute-heavy prefill to memory-bound decode, the energy-optimal SM/frequency setting can change during execution. Each GPU runs a local scheduler that monitors the live batch mix and dynamically adjusts SM partitioning and GPU frequency to track these phase changes without violating SLOs.

\noindent $\bullet$ \textbf{Cluster-Level Consolidation: Active Scale-In (\S\ref{ss:design_LLM_replacement}).}
To reduce cluster-wide static power (e.g., HBM) under low load, \systemname periodically re-evaluates placement and consolidates workloads onto fewer GPUs, deactivating underutilized devices when possible while preserving frequency affinity. 
During migration, \systemname~leverages the graph algorithm to construct the transmission plan and selectively keeps the near-completion requests locally instead of transferring.

\vspace{-2mm}
\subsection{One-Time Offline Profiling}
\label{ss:design_offline_profiling}

Our offline profiling consists of Look-Up Table construction and output token length predictions.


\subsubsection{Look-Up Table Construction}
\label{sss:lut}

\begin{figure}[t]     
  \centering
  \includegraphics[width=\linewidth]{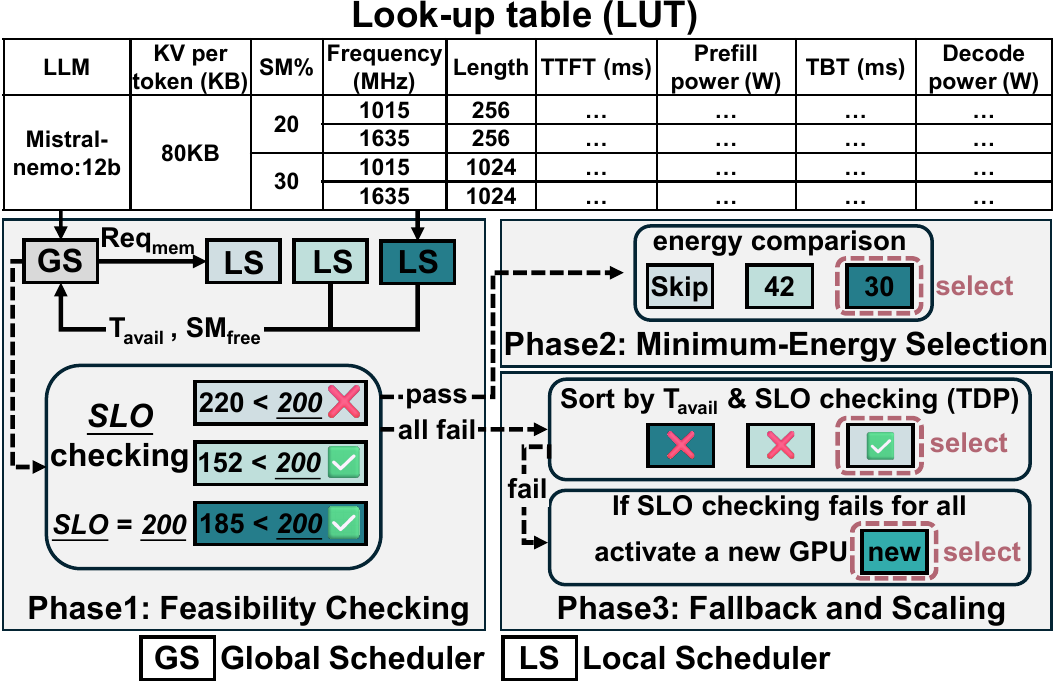}
  \vspace{-8mm}
  \caption{Dispatching flow of the global scheduler.}
  \label{fig:global_scheduler}
  \vspace{-5mm}             
\end{figure}

Prior to runtime deployment, \systemname constructs a high-fidelity look-up table that records the inference latency and power metrics of each LLM under different request intensity and GPU hardware resource configurations, as shown in Figure~\ref{fig:global_scheduler}. 
We perform a comprehensive grid search across three primary dimensions: GPU clock frequency, Streaming Multiprocessor (SM) allocation, and input sequence length.

\noindent\textbf{Profiling Grid and Bounds.} We sweep GPU clock frequencies from 795 MHz to 1635 MHz with a stride of 90 MHz. 
The lower bound of 795 MHz is chosen based on findings from EVeREST~\cite{everest}, which demonstrate that reducing frequency below this threshold yields negligible voltage reduction, thereby halting energy savings.
The upper bound corresponds to TDP for specific GPUs (e.g., 1635 MHz for H100), consistent with prior energy-aware frameworks~\cite{dynamollm,everest}. 

A small frequency step (e.g., 15~MHz) enables fine-grained control over energy consumption. 
However, adjusting the frequency by 15~MHz in either direction around the optimum changes total energy consumption by less than 1\%, as shown in Figure~\ref{fig:motivation_energu_frequency_pattern}.
In contrast, a coarser 90~MHz step yields an energy difference of approximately 5\% but reduces profiling overhead by roughly $6\times$.
This coarser step size is also consistent with prior work~\cite{dynamollm,yu2025voltanallm}.

We also vary SM allocation from 10\% to 100\% with a stride of 10\%. 
To ensure our profiling captures representative production demands, we define the input sequence range based on real-world LLM workload traces from Azure~\cite{dynamollm}, sweeping from 256 to 8192 tokens (same as DynamoLLM~\cite{dynamollm}) to encompass the vast majority of standard inference requests.

\noindent\textbf{Metric Collection and Interpolation.} For each configuration tuple (Frequency, SM\%, Input Length), we record three critical metrics: Time To First Token (TTFT), Time Between Tokens (TBT), and average power consumption. 
We also quantify the one-time latency of loading model weights into GPU memory. 
To estimate performance for input lengths not explicitly covered by our grid, we employ polynomial regression, a standard technique for continuous resource modeling in LLM serving~\cite{dynamollm,huawei-dvfs,hedrarag}. 

\noindent\textbf{Profiling Overhead and Reusability.} Note that offline profiling is a one-time, low-overhead setup step. In our implementation, the full profiling procedure completes in under 40 minutes on an NVIDIA H100. Although profiling is hardware- and model-specific, it is required only once per GPU type and model. Moreover, serverless providers typically offer a small, fixed set of GPU SKUs and model types; therefore, the total number of distinct profiles needed in practice is limited. 




\subsubsection{Output Token Length Prediction.} 
\label{sss:token_length_predictor}

To accurately estimate total inference latency and determine if a specific GPU frequency can satisfy a request's SLO, \systemname should account for the duration of each task's prefill and decode phase.
\begin{wrapfigure}{r}{0.25\textwidth}     
\vspace{-4mm}
  \centering
  \includegraphics[width=1\linewidth]{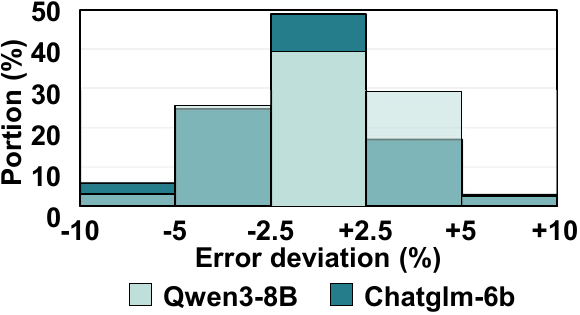}
  \vspace{-6mm}
  \caption{Distribution of relative error percentages, categorized by deviation ranges.}
  \label{fig:prediction_error_distribution}
  \vspace{-4mm}             
\end{wrapfigure}
Prior works~\cite{dynamollm,output_length_pred_s3,output_length_pred_sequence_scheduling,output_length_pred_ssjf} have shown that the output token length is highly predictable.
We thus borrow the same approach to predict the output token length using a high-accuracy output length predictor~\cite{dynamollm,output_length_pred_sequence_scheduling}.

Figure~\ref{fig:prediction_error_distribution} illustrates the distribution of prediction errors of two LLMs under real-world LLM workload traces~\cite{dynamollm,lmsyschat_trace}. 
As observed, the predictor exhibits a tight error bound, with the vast majority of predictions deviating by less than 5\% from the ground truth (only 2\% of predictions have an error > +5\%). 
Suggested by these results, we provide a modest extra 5\% memory buffer, which sufficiently accommodates this variance without incurring significant over-provisioning waste.
With this, the global scheduler can pre-calculate the expected latency for any given frequency-SM configuration, ensuring that dispatching decisions are both energy-efficient and SLO-compliant.


\vspace{-3mm}
\subsection{\textit{The Global Scheduler}: Light-Weight, Frequency-Aware Request Dispatching}
\label{ss:design_freq_aware_dispatching}


The energy-optimal GPU frequency depends strongly on the request's workload intensity: running at a mismatched frequency can substantially increase energy consumption, while running too slowly risks SLO violations. To address this, we propose a frequency-aware dispatcher that routes each request to a GPU whose \emph{current} clock frequency best matches the request’s workload, while still meeting the SLO. 
As illustrated in Figure~\ref{fig:global_scheduler}, the dispatcher works in three phases: (i) feasibility checking, (ii) GPU selection, and (iii) resource scaling. We elaborate on each phase below.

\noindent $\bullet$ \textbf{Phase One: Pooling-based Feasibility Checking.} When a request arrives, the global dispatcher first predicts its output token length and adds a 5\% safety margin to account for estimation error (\S\ref{sss:token_length_predictor}). 
With request input length $Req_{in}$ and this padded length $Req_{out}$, it derives the request’s memory demand, denoted as $Req_{mem}$.

Next, the dispatcher polls all local schedulers to determine which GPUs can accommodate the request. Each local scheduler maintains the state of its running and queued requests and uses the offline-profiling look-up table (\S\ref{ss:design_offline_profiling}) to translate these queued workloads into a time-varying estimate of SM and memory utilization. From this local view, the scheduler computes (1) the \textit{Earliest Available Timestamp} $T_{avail}$ at which at least $Req_{mem}$ memory can be allocated, and (2) the number of free SMs available at that time, $SM_{free}$. It then returns $(T_{avail}, SM_{free})$ to the global dispatcher.

Finally, the global dispatcher performs an SLO feasibility check for each GPU. A GPU is considered feasible only if a request starting at $T_{avail}$, with $SM_{free}$ SMs, can complete before its SLO deadline under the GPU’s current clock frequency $F_{curr}$. This phase ensures that subsequent energy-oriented decisions are made only among candidates that do not compromise SLO compliance.
As shown in Phase 1 of Figure~\ref{fig:global_scheduler}, only GPUs with estimated timestamp < 200 (i.e., SLO target) are feasible for the next phase.


\noindent  $\bullet$ \textbf{Phase Two: Energy-Minimal GPU Selection.} 
Given the feasible set of candidates $\mathcal{C}$, the global dispatcher chooses the target GPU that minimizes the request’s expected energy consumption at the GPU’s current operating frequency. For each candidate $GPU_k \in \mathcal{C}$ with current frequency $f_k$ and SM availability $S_k$, the dispatcher estimates the energy to serve the request as:
\vspace{-2mm}
\begin{equation}E^{k} = \text{Power}(f_k) \times \text{Latency}(f_k, S_k, Req_{in}, Req_{out})\end{equation}
where $\text{Power}(\cdot)$ and $\text{Latency}(\cdot)$ are obtained from the look-up table. 
The scheduler dispatches the request to the GPU $k^*$ that minimizes the energy cost:
\vspace{-2mm}
\begin{equation}
k^* = \operatorname*{argmin}_{k \in \mathcal{C}} \left( E^{k} \right)
\end{equation}

\begin{figure}[t]     
  \centering
  \includegraphics[width=\linewidth]{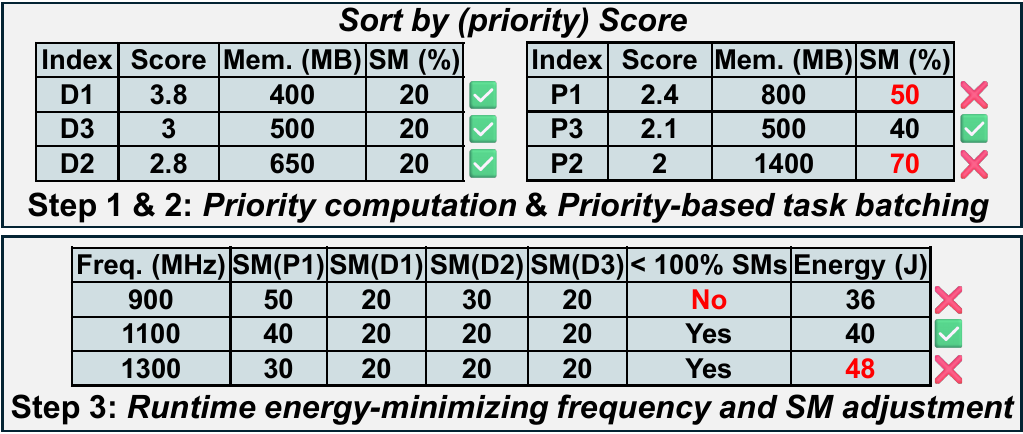}
  \vspace{-8mm}
  \caption{Working flow of the local scheduler.}
\label{fig:local_scheduler}
  \vspace{-5mm}            
\end{figure}

In Phase 2 of Figure~\ref{fig:global_scheduler}, the GPU with lower energy consumption (i.e., 30 Joules) is selected to process the request.
This objective implicitly favors frequency alignment. Because energy exhibits a convex dependence on frequency (\S\ref{sec:motivation_u_shape}), the estimated energy $E^{k}$ increases when a candidate’s current frequency $f_k$ is far above or below the request’s energy-optimal operating point (i.e., frequency sweet spot), making such mismatches less likely to be selected.

It should be noticed that when multiple candidates have near-identical energy (e.g., within 2\%), \systemname uses a stability-oriented tie-breaker: it first selects the GPU with the most free memory to reduce out-of-memory risk, and then the GPU with the largest $SM_{free}$ to improve throughput.

\noindent $\bullet$ \textbf{Phase Three: Fallback and Scaling.} If the set of candidates $\mathcal{C}$ is empty, which means no GPU can meet the SLO at its current frequency, the global dispatcher relaxes the frequency constraint. 
It orders GPUs by the earliest available timestamp and evaluates feasibility assuming each GPU boosts to its TDP-limited maximum frequency ($F_{max}=1635$ MHz).
The request is dispatched to the first GPU capable of meeting the SLO under $F_{max}$.
If the deadline remains infeasible even at $F_{max}$, \systemname triggers a scale-out action to activate a new GPU, as exhibited in Phase 3 of Figure~\ref{fig:global_scheduler}. 
For this newly activated GPU, the global scheduler queries the LUT to identify the optimal frequency, minimizing energy usage without violating SLOs. This scheduling process has negligible overheads (< 1 ms for 10,000 GPUs).
\subsection{\textit{The Local Scheduler}: Joint Frequency-\ SM Adaptation for Energy Efficiency}
\label{ss:design_fine-grained_adaptation}

The local scheduler deployed on each GPU decides which prefill and/or decode tasks to run next and adjusts SM partitions and GPU clock for those tasks, performing fine-grained resource management to optimize energy efficiency. Figure~\ref{fig:local_scheduler} illustrates the working flow of the local scheduler.

Motivated by prior LLM serving work on stall-free scheduling~\cite{muxserve,gao2025weaver,FlashGen,patke2025queuemanagementsloorientedlarge,choi2025elisefficientllmiterative}, the local scheduler allows \emph{out-of-order interleaving} between prefill and decode so that a request's short decode steps can still make progress even when another request's long prompts are present. The key challenge is that these two task types operate on very different time scales: decode deadlines are at the millisecond level, whereas prefill deadlines are typically at the second level.


\noindent\textbf{Unified Priority Score.} To compare these heterogeneous tasks directly, we normalize them using a single urgency metric: the \textit{Slack Consumption Ratio} ($\text{Score}_{task}$). It measures the fraction of the remaining time  budget a task would consume if scheduled next; higher values indicate higher urgency
:
\begin{equation}\label{eq:unified_score}\text{Score}_{task} = \frac{\text{Latency}(task)}{\text{Slack}_{task} - \beta \cdot \text{Age}_{task}}\end{equation}
In above equation, $\text{Latency}(task)$ is the profiled execution time of the scheduling unit (one decode step for decode tasks, and a prefill unit for prefill tasks), and $\text{Slack}_{task}$ is the remaining time budget to the relevant SLO deadline (i.e., TBT for decode; TTFT for prefill). $\text{Age}_{task}$ is the time a task has waited since becoming runnable. We set $\beta=0$ for decode tasks (they are continuously active and repeatedly re-evaluated), and use $\beta=1$ for prefills so that waiting gradually reduces effective slack, preventing starvation.


\noindent\textbf{Task Batching.} 
Using this unified score, the scheduler maintains a single priority queue containing both runnable decodes and pending prefills. At each epoch, it constructs the next batch by selecting tasks in descending $\text{Score}_{task}$ order subject to the GPU's instantaneous memory capacity and the chosen SM partition.
This resource-aware selection is illustrated in Figure~\ref{fig:local_scheduler}. The scheduler iteratively fills the batch. However, if the remaining resources (e.g., 40\% SM capacity) are insufficient for the highest-priority task (P1), the scheduler bypasses it to dispatch a fitting lower-priority task (P3). 
This strategy mitigates head-of-line blocking and maximizes resource occupancy. 
Consequently, the policy dynamically balances trade-offs: it prioritizes decodes when TBT slack is tight, yet elevates prefills as their TTFT deadlines approach, ensuring SLO compliance across both phases.


\noindent\textbf{Runtime Resource Allocation.}
Given a candidate batch, the scheduler jointly chooses GPU clock frequency and SM partitioning to minimize energy while meeting the batch's SLO constraints. 
Similar in spirit to DVFS-based energy optimization~\cite{you2023zeus}, we exploit the fact that valid GPU clock settings are discrete and few (typically $<10$ levels~\cite{dynamollm,everest}). 
Rather than solving a complex continuous optimization, we perform a lightweight \emph{frequency sweep}, exhibited in Step 3 of Figure~\ref{fig:local_scheduler}:
for each candidate frequency $f$, we use the offline look-up table to (1) compute, for each task in the batch, the minimum SM allocation required to satisfy its deadline at frequency $f$, and (2) check whether the aggregate SM demand fits within the device budget (i.e., < 100\% SMs). For feasible configurations, we estimate total energy using the profiled power and latency models, and select the frequency/partition pair with the lowest predicted energy. Because the sweep iterates over a small constant-sized frequency set and uses only a look-up table, its overhead is negligible (i.e., 0.1 ms), scaling as $O(|\mathcal{F}| \cdot N)$ for $N$ tasks and $|\mathcal{F}|$ frequency levels.

\noindent\textbf{Handle mis-predictions}. In rare instances where the global scheduler underestimates output length, a request may exceed its allocated memory budget, triggering an Out-of-Memory (OOM) error. To resolve this, the local scheduler immediately assigns the highest priority to the mispredicted request.
If memory remains insufficient to complete this request, the local scheduler evicts the running request that incurs the minimum recomputation overhead to free up space.

\vspace{-3mm}
\subsection{Energy-Aware Workload Replacement}
\label{ss:design_LLM_replacement}

To further reduce the cluster-wide energy overhead, \systemname periodically consolidates workload by migrating LLM workloads off lightly loaded GPUs onto more utilized ones and deactivates the freed GPUs.
To make consolidation energy-aware, we introduce a metric called \emph{Energy Waste Ratio (EWR)} (\S\ref{ss:ewr_defination}) and then design an EWR-aware reconfiguration algorithm that consolidates aggressively while minimizing the energy penalty from frequency misalignment (\S\ref{ss:ewr_scheduling}).

\begin{figure}[t]     
  \centering
  \includegraphics[width=\linewidth]{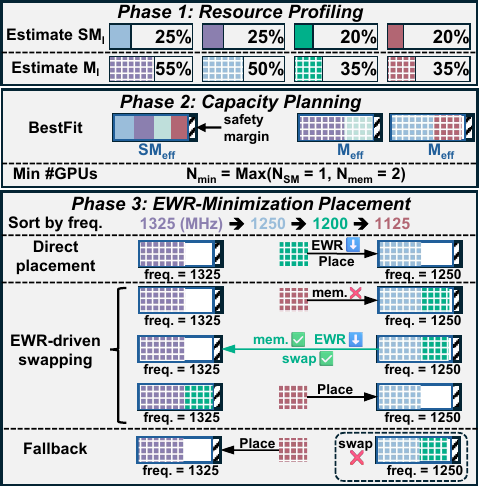}
  \vspace{-8mm}
  \caption{EWR-aware LLM placement reconfiguration.}
  \label{fig:replacement}
  \vspace{-2mm}             
\end{figure}

\subsubsection{Energy Waste Ratio (EWR)}
\label{ss:ewr_defination}
We quantify consolid-\
ation-induced energy waste as the gap between the power implied by a GPU's \emph{actual} operating frequency after co-location and the power each LLM would ideally consume at its own energy-optimal frequency. 
Since GPU dynamic power scales approximately cubically with frequency ($P \propto f^3$)~\cite{wierman2009power,han2025joint,liu2025greenllm,power_voltage_frequency}, the cluster-wide energy waste can be expressed (up to a constant factor) as:
\vspace{-1mm}
\begin{equation}\label{eq:energy_waste}E_{\text{waste}} = \sum_{g \in \text{GPUs}} \sum_{l \in g} t_l \cdot (F^3_{g} - F^3_l)\end{equation}
where $F_g$ represents the operating frequency of GPU $g$ (determined by the highest frequency of co-located LLMs); $F_l$ is the desired energy-optimal frequency of LLM $l$; $t_l$ is the model execution time. 
As a reference, we define the idealized energy if every LLM ran at its preferred frequency:
\begin{equation}
\label{eq:energy_optimal}
E_{\text{optimal}} = \sum_{g \in \text{GPUs}} \sum_{l \in g} t_l \cdot F^{3}_{l}
\end{equation}

We then derive the \emph{Energy Waste Ratio} as a normalized inefficiency metric:
\vspace{-2mm}
\begin{equation}
\label{eq:energy_waste_ratio}
\text{EWR} = \frac{E_{\text{waste}}}{E_{\text{optimal}}}
\end{equation}

$\text{EWR}=0$ indicates perfect frequency alignment (no waste), while larger values imply increasing energy loss due to consolidation-induced frequency skew.

\subsubsection{EWR-Aware Workload Replacement}
\label{ss:ewr_scheduling}
Figure~\ref{fig:replacement} depicts the three phases to minimize active GPUs while maintaining a low EWR, the full algorithm is stated in  Algo.\ref{alg:placement_reconfig}.

\begin{algorithm}[th]
\footnotesize
\caption{EWR-Aware Placement Reconfiguration}
\label{alg:placement_reconfig}
\SetAlgoLined
\DontPrintSemicolon

\SetKwInOut{Input}{Input}
\SetKwInOut{Output}{Output}
\SetKwFunction{EstSM}{EstimateSM}
\SetKwFunction{EstMem}{EstimateMem}
\SetKwFunction{CalcFreq}{CalcFreq}
\SetKwFunction{CalcEWR}{CalcEWR}
\SetKwFunction{Fits}{Fits}
\SetKwFunction{Sort}{Sort}

\Input{List of active LLMs $\mathcal{L}$; GPU Constraints ($SM_{eff}, M_{eff}$)}
\Output{Map of LLMs to GPUs}

\tcp{Phase 1: Resource Profiling}
\ForEach{LLM $l \in \mathcal{L}$}{
    $l.S \leftarrow$ \EstSM{model, avg\_in, avg\_out}\;
    $l.M \leftarrow$ \EstMem{model\_size, max\_kv}\;
    $l.F \leftarrow$ \CalcFreq{model, $l.S$, avg\_in, avg\_out}\;
}

\tcp{Phase 2: Capacity Planning}
$N_{sm} \leftarrow BestFitDecreasing(\mathcal{L}.S,\ SM_{eff})$\;
$N_{mem} \leftarrow BestFitDecreasing(\mathcal{L}.M,\ M_{eff})$\;
$N_{min} \leftarrow \max(N_{sm},\ N_{mem})$\;
Initialize active GPUs set $\mathcal{G}$ with count $N_{min}$\;

\tcp{Phase 3: EWR Minimization Loop}
\Sort{$\mathcal{L}$ by $l.F$}\tcp*[r]{Group by Frequency}
\ForEach{LLM $l \in \mathcal{L}$}{
    $g_{target} \leftarrow \argmin_{g \in \mathcal{G}} |g.freq - l.F|$\;
    $g_{cand} \leftarrow \argmin_{g \in \mathcal{G} \setminus g_{target}} |g.freq - l.F|$\;
    
    \uIf{\Fits($g_{target}, l$)}{
        Assign $l$ to $g_{target}$\;
    }
    \uElse{
        $l_{out} \leftarrow$ Incumbent on $g_{target}$ with lowest $F$\;
        \tcp{Simulate Swap to check EWR reduction}
        $E_{curr} \leftarrow$ \CalcEWR($g_{target} \cup \{l\}, g_{cand}$)\;
        $E_{swap} \leftarrow$ \CalcEWR($g_{target} \setminus \{l_{out}\} \cup \{l\}, g_{cand} \cup \{l_{out}\}$)\;
        
        \uIf{\Fits($g_{cand}, l_{out}$) \textbf{and} \Fits($g_{target} \setminus l_{out}, l$) \textbf{and} $E_{swap} < E_{curr}$}{
            Move $l_{out}$ to $g_{cand}$; Assign $l$ to $g_{target}$\;
        }
        \uElse{
             \uIf{\Fits($g_{cand}, l$)}{
                 Assign $l$ to $g_{cand}$\;
             }
             \Else{
                 Activate new GPU $g_{new}$ and assign $l$\;
                 $\mathcal{G} \leftarrow \mathcal{G} \cup \{g_{new}\}$\;
             }
        }
    }
    Update $g.SM, g.M, g.freq$ for all affected GPUs\;
}
\end{algorithm}

\noindent\textbf{Phase One: Profiling \& Resource Estimation.} 
For each active LLM $l$, the local scheduler calculates the preferred configuration ($SM_l, F_l$) that minimizes energy while meeting SLOs, based on the average input/output lengths in the past time window (i.e., $\Delta$ = 5 minutes).
The local scheduler queries the Look-Up Table (LUT) to identify the optimal independent configurations for the prefill and decode phases---denoted as ($SM_p, F_p$) and ($SM_d, F_d$)---that minimize energy consumption while satisfying SLOs. The LUT also provides the expected power consumption ($P_p, P_d$) and execution duration ($T_p, T_d$) for each phase under these settings.
To derive a unified configuration ($SM_l$), we compute the time-weighted average of the required SMs, ensuring the allocation reflects the phase occupying the majority of the execution time:
\begin{equation}
\label{eq:preferred_sm}SM_l = \frac{SM_p \cdot T_p + SM_d \cdot T_d}{T_p + T_d}
\end{equation}
Simultaneously, it estimates the memory requirement, $M_l$, using Little's Law~\cite{newell2013applications} to avoid static over-provisioning. Specifically, the local scheduler calculates the expected request concurrency using the formula $N_{conc} = \lambda \cdot \mathcal{W}$, where $\lambda$ is the measured request arrival rate and $\mathcal{W}$ is the average processing latency given the GPU configuration ($SM_l, F_l$). The total memory pressure is then derived by summing the static model weights and the KV cache required for $N_{conc}$ active sequences, expressed as: $M_l = \text{Weights} + N_{conc} \cdot \text{Size}_{KV}$.

\noindent\textbf{Phase Two: Global Capacity Planning.} Before choosing specific GPUs, the global scheduler computes the minimum number of GPUs ($N_{gpu}$) required to serve the aggregate workload.  
However,  na\"ive packing strategies are susceptible to resource thrashing, where minor fluctuations in LLM workloads (e.g., request rate) trigger a scale-out shortly after consolidation, negating energy savings and incurring cold-start penalties.
To mitigate resource thrashing, we reserve a safety margin (e.g., $\delta=5\%$ in this work) for both SMs and memory on active GPUs. We define the effective capacity as $SM_{eff}=(1-\delta)\cdot SM_{cap}$ and $M_{eff}=(1-\delta)\cdot M_{cap}$.
Then, a bin-packing lower bound is solved for $SM_l$ and $M_l$ with respect to $SM_{eff}$ and $M_{eff}$, and the maximum is used as $N_{gpu}$.
This step identifies the smallest feasible active GPU pool for the current workload while retaining headroom to absorb minor workload fluctuations without triggering scale-out.

\begin{table}[t]
    \centering
    \caption{TTFT/TBT SLOs from DynamoLLM~\cite{dynamollm}.} 
    
    \label{tab:slo_setup}
    \vspace{-4mm}
    \begin{tabular}{cccccc}
        \toprule
         & & Input & Output & TTFT SLO & TBT SLO \\ 
        \midrule
        Short & S & $<256$ & $<100$ & 250 ms & 100 ms \\ 
        Medium & M & $<1024$ & $<350$ & 400 ms & 100 ms \\ 
        Long & L & $\le 8192$ & $\ge 350$ & 2000 ms & 100 ms \\ 
        \bottomrule
    \end{tabular}
    \vspace{-8mm}
\end{table}

\noindent\textbf{Phase Three: Workload Replacement.} Given $N_{\text{gpu}}$ active GPUs, the scheduler places LLMs to reduce frequency skewness by grouping models with similar preferred frequencies. It sorts all LLMs by $F_l$ and assigns them in order. 
For each LLM, it identifies a \textit{Target GPU} (the active GPU with the closest operating frequency) and a \textit{Candidate GPU} (the second closest) based on the following rules:

$\bullet$ \textbf{Direct Placement:} If the \textit{Target GPU} satisfies resource constraints ($\geq SM_l, M_l$), the LLM is assigned here. This minimizes $F^3_{g} - F^3_l$, keeping EWR low.


$\bullet$ \textbf{EWR-Driven Swapping:} If the \textit{Target GPU} is saturated, the scheduler attempts to migrate an "outlier" incumbent (the resident LLM with the lowest $F_l$) to the \textit{Candidate GPU}. If the swap is resource-valid and reduces the total $E_{\text{waste}}$, the outlier is moved to accommodate the new LLM. 


$\bullet$ \textbf{Fallback:} If prior actions fail, the scheduler iteratively checks the next closest GPUs in the sorted list and assigns the LLM to the first GPU found with sufficient resources.
\noindent\textbf{A Running Example of Phase Three}. 
As shown in Figure~\ref{fig:replacement}, three LLMs in purple, blue, and green are successfully placed onto their preferred nodes without contention (Direct Placement).
Next (EWR-driven swapping), when trying to place the red LLM on the second GPU, which gives the closest frequency, the memory is insufficient to hold the red LLM. The green LLM is then moved from the second GPU to the first, due to sufficient resources on the first GPU for the green LLM and the reduced overall EWR, then the red LLM is placed on the second GPU.
If the prior two actions fail to deploy the red LLM, the scheduler iteratively checks the next closest GPUs in the sorted list. The LLM is placed on the first active GPU found with sufficient resources. In this case, the red LLM is placed on the first GPU, despite encountering a huge frequency deviation.

\noindent\textbf{When to trigger workload replacement?}
We trigger a replacement check when an instance’s keep-alive timer expires, indicating an LLM has become idle and can be evicted without disruption. A replacement is performed only if capacity planning (with a safety margin for small fluctuations) shows the workload can be served with fewer GPUs ($N_{\text{gpu}} < N_{\text{active}}$), enabling scale-in with low control-plane overhead.

\noindent
\textbf{Mitigate resource fragmentation. }
\systemname~reclaims resource fragmentation (e.g., SM, memory) at both cluster and device levels.
At the cluster level, \systemname~periodically tries to consolidate workloads onto fewer GPUs without violating SLOs (\S3.5). If the fragmented capacity is sufficient to release a GPU, \systemname~migrates and redistributes workloads and then deactivates the freed device.
At the device level, \systemname's local scheduler dynamically adjusts SM allocations for each task batch (\S3.4). It continues adding runnable tasks while their SM, memory,and SLO constraints can be satisfied. Thus, small residual fragments remain available to subsequent tasks rather than being permanently stranded.

\begin{figure*}[t]     
  \centering
  \includegraphics[width=\linewidth]{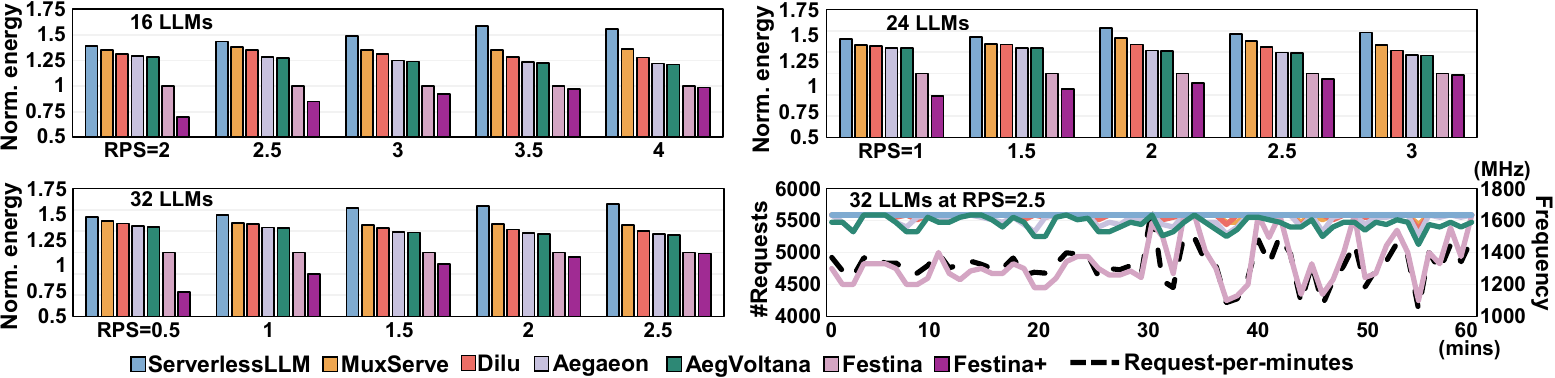}
  \vspace{-8mm}
  \caption{Normalized energy consumption relative to \systemname under varying RPS loads and numbers of LLMs, alongside a runtime trace of GPU frequency and request arrival rate for the 32-LLM scenario at 2.5 RPS.}
  \label{fig:main_energy}
\end{figure*}

\vspace{-3mm}
\section{Evaluation}
\label{s:evaluation}

\systemname is implemented in approximately 2,800 lines of Python and CUDA/C++ code. 
The scheduler is built using Python's asyncio~\cite{asyncio} library to enable high-concurrency instance orchestration. 
The backend is built upon vLLM~\cite{vLLM_2026} to benefit from state-of-the-art optimizations such as FlashAttention~\cite{dao2023flashattention} and PagedAttention~\cite{kwon2023efficient}. 
We utilize pyNVML~\cite{pynvml} to adjust GPU frequency with minimal runtime overhead.

We run experiments to answer the following questions:

(1): {\it How does \systemname balance SLO attainment with energy efficiency compared to SOTA serverless LLM systems?}

(2): {\it Can \systemname maintain performance across different workload intensities and types?}


(3): {\it What is the contribution of each design module?}

(4): {\it How sensitive is the system to configuration parameters like safety margins and disaggregated setups?}


\vspace{-3mm}
\subsection{Experimental Setup}
\label{ss:exp_setup}

\noindent\textbf{Testbed.} We conduct experiments on a serverless testbed consisting of eight NVIDIA H100 GPUs, interconnected via NVLink. All GPUs support Multi-Process Service (MPS) for SM partitioning.

\noindent\textbf{Models, Datasets, and Workloads.} Our evaluation focuses on LLMs that fit on a single GPU, since \systemname targets serverless GPU sharing where each GPU can co-host multiple models. 
With that, we select representative LLMs from various families, including Qwen, Llama, Yi, and Mistral. Models range from 3B to 14B parameters, consistent with prior studies~\cite{muxserve, aegaeon}.
We evaluate our system using the ShareGPT dataset.
Following prior works~\cite{aegaeon,muxserve,fu2024serverlessllm,dilu, servegen}, we generate workloads with a scaled Poisson process
and random sampling from the dataset.
To stress-test \systemname under prefill-heavy and decode-heavy scenarios, we also use two variants from Aegaeon~\cite{aegaeon}: ShareGPT-ix2 (2$\times$ input length) and ShareGPT-ox2 (2$\times$ output length).

\noindent\textbf{Metrics.} We measure energy consumption and SLO attainment, adopting SLO constraints from DynamoLLM~\cite{dynamollm} (Table~\ref{tab:slo_setup}). Requests exceeding 8192 tokens are excluded. Energy is profiled from the first request arrival to the final completion.
We report Normalized Energy Consumption, calculated as the total energy consumed by a baseline system divided by that of \systemname for the same workload. A value greater than 1.0 indicates higher energy usage than \systemname.

\noindent\textbf{Baselines.} We compare \systemname against four open-sourced SOTA serving systems representing the latest works in serverless auto-scaling, GPU multiplexing, and PD-disaggregation. 
\noindent$\bullet$ \textbf{ServerlessLLM}~\cite{fu2024serverlessllm}: A specialized serverless framework designed for fast auto-scaling. It minimizes cold-start latency through rapid checkpoint loading and implements efficient live migration mechanisms.

\noindent$\bullet$ \textbf{MuxServe}~\cite{muxserve}: A static multiplexing system that optimizes SM partitioning and model placement offline. To ensure a fair comparison with our dynamic approach, we extended MuxServe to support co-locating more than two models per GPU and added a keep-alive mechanism to offload idle models to host memory.

\noindent$\bullet$ \textbf{Dilu}~\cite{dilu}: A dynamic serving system that optimizes cluster utilization by scheduling models according to real-time demand. It employs on-the-fly SM re-partitioning to minimize latency for co-located workloads.

\noindent$\bullet$ \textbf{Aegaeon}~\cite{aegaeon}: The state-of-the-art system for prefill-decode disaggregated serverless serving. Aegaeon decouples prefill and decode phases onto separate nodes, using fine-grained KV cache management to reduce data transmission overhead and MPS-based sharing to maximize GPU occupancy.

\noindent$\bullet$ \textbf{AegVoltana}: We augment Aegaeon with VoltanaLLM~\cite{yu2025voltanallm} to construct a DVFS-aware baseline. 

\noindent$\bullet$ \textbf{\textsf{Festina+}\xspace}: \textsf{Festina+}\xspace simulates the effects of dynamic HBM frequency scaling (a feature not yet supported on NVIDIA H100) during idle periods. We include this to establish an upper bound on the potential energy savings our system can achieve through memory frequency adjustments after scale-in LLM placement reconfiguration in Section~\ref{ss:design_LLM_replacement}.




\begin{figure*}[t]     
  \centering
  \includegraphics[width=\linewidth]{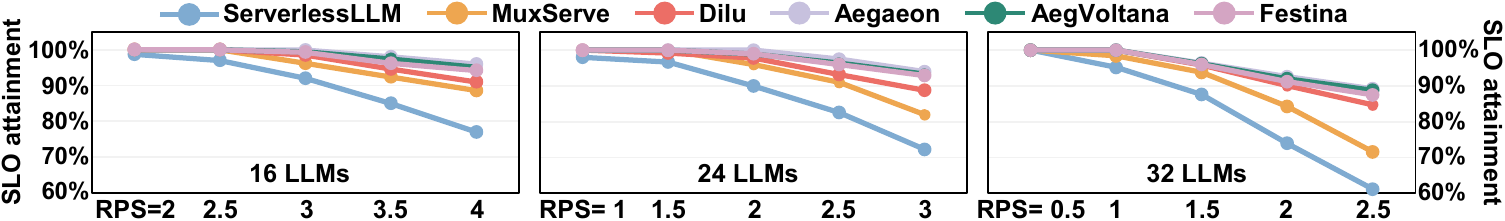}
  \vspace{-7mm}
  \caption{SLO attainment under various average request-per-second (RPS).}
  \label{fig:main_slo}
  \vspace{-3mm}         
\end{figure*}

\vspace{-2mm}
\subsection{Head-to-head Comparison with SOTAs}
\label{ss:main_results}

\vspace{-1mm}
\subsubsection{Energy Consumption} 
\begin{figure}[]     
  \centering
  \includegraphics[width=\columnwidth]{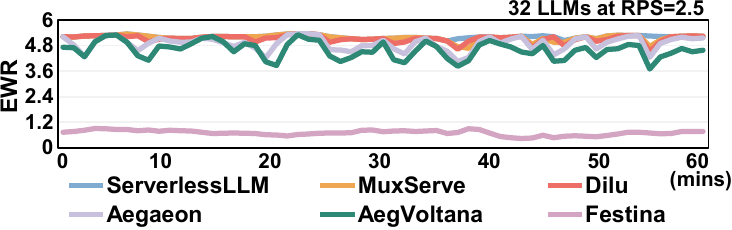}
  \vspace{-8mm}
  \caption{EWR over time when serving 32 LLMs at RPS=2.5. \systemname~averages 0.71, lower than baselines by over 7$\times$.}
  \label{fig:ewr_overtime}
  \vspace{-8mm}       
\end{figure}

Figure~\ref{fig:main_energy} shows the normalized energy consumption under various average request-per-second (RPS) and numbers of LLMs settings, where a value greater than 1.0 indicates higher energy usage than \systemname.
\systemname consistently uses less energy compared to all baselines, where ServerlessLLM, MuxServe, Dilu, Aegaeon, and AegVoltana consume up to 1.56$\times$, 1.37$\times$, 1.36$\times$, 1.31$\times$, and 1.3$\times$ energy than \systemname, respectively.
The reason behind this energy gap is two-fold. (i): the global scheduler of \systemname will dispatch requests to GPUs based not only on SLO constraints but also on frequency preference. This means co-located LLMs will process their requests at a frequency that won't hugely deviate from their own optimal frequency, while other baselines don't employ such a frequency-aware dispatching mechanism.
(ii): these baselines target higher SLO attainment, which means they adjust resource allocation to speed up inference latency without optimizing energy efficiency. 
In contrast, the local scheduler of \systemname dynamically reconfigures not only GPU frequency but also SM allocation to meet SLO constraints and minimize energy consumption.
Although AegVoltana adjusts per-GPU frequency based on SLO slack, it lacks frequency-aware request dispatching and model placement. 
As a result, models with mismatched frequency preferences are colocated on the same GPU and forced to operate at a frequency dictated by their neighbors rather than their own energy-optimal point. 
\systemname~mitigates these mismatches by reducing EWR during periodic consolidation and its frequency-aware request dispatching mechanism. 
Figure~\ref{fig:ewr_overtime} depicts the overtime EWR for \systemname~and baselines when serving 32 LLMs at RPS=2.5.
As one can observe, \systemname~achieves an EWR of 0.71, compared with 4.96 for AegVoltana and 5.18 for Aegaeon, where \systemname~reduces EWR by more than 85\%. 
This EWR reduction achieved by \systemname~and the limited energy reduction from DVFS alone in AegVoltana underscores the importance of jointly SLO- and frequency-aware scheduling, which is the core capability of \systemname.

We further plot the GPU frequency of all systems with respect to the workload variation in Figure~\ref{fig:main_energy}.
As shown, the GPU frequency in \systemname closely follows the workload pattern up and down, whereas all other baseline systems keep the GPU frequency at a consistently high level across the entire one-hour observation window.
In summary, \systemname can save at least 22\% energy under a heavy-loaded scenario. If the vendors provide the function to adjust HBM frequency, our system (\textsf{Festina+}\xspace) can further save up to 48\% energy.

\begin{figure*}[t]     
  \centering
  \includegraphics[width=\linewidth]{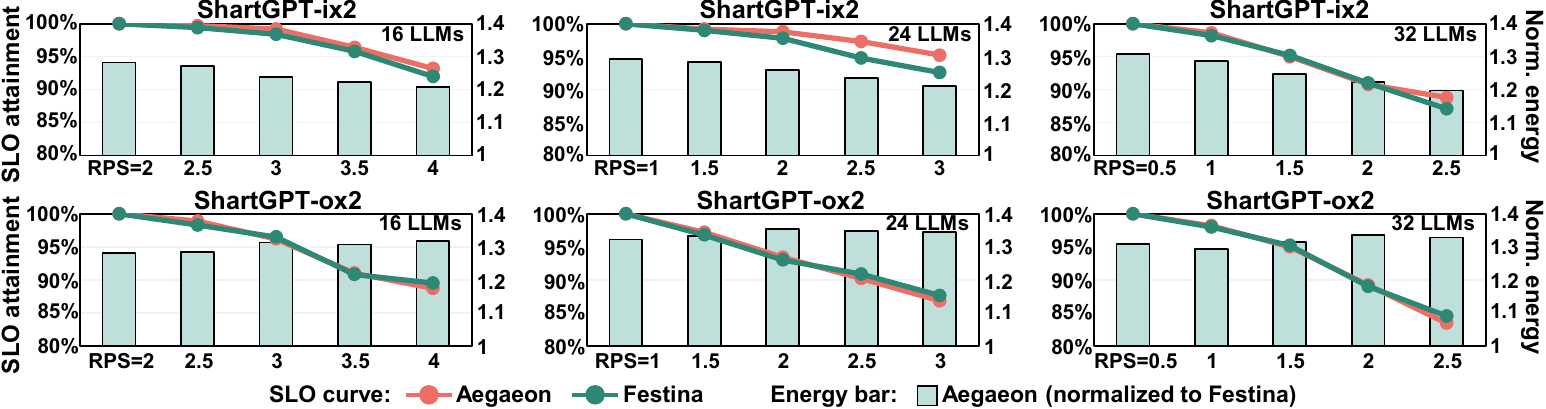}
  \vspace{-9mm}
  \caption{SLO attainment and normalized energy consumption under various RPS and \# of LLMs.}
  \label{fig:input_length_x2}
  \vspace{-5mm}          
\end{figure*}

\vspace{-12mm}
\subsubsection{SLO Attainment}
Next, we measure the latency SLO attainment of all systems under the same setup as the previous experiment. As shown in Figure~\ref{fig:main_slo}, \systemname outperforms ServerlessLLM, MuxServe, and Dilu by up to 27\%, 16\%, and 4\%, respectively. Also, \systemname exhibits comparable performance to Aegaeon and AegVoltana, with gaps of less than 2\% and 1\%, respectively.

Specifically, \systemname leverages MPS to allow multiple LLMs to share the same GPU and to process requests concurrently, which improves device utilization and shortens the overall execution time. However, ServerlessLLM utilizes a temporal sharing manner that each LLM occupies the entire GPU in turn, which under-utilizes GPUs and prolongs the waiting time of co-located LLMs. This leads to significant SLO violation, especially when request-per-second becomes high.
MuxServe relies on static offline partitioning, preventing models from utilizing idle SMs beyond their pre-assigned quota. 
Dilu, despite supporting dynamic resizing, employs a First-Come-First-Served (FCFS) policy that exacerbates Head-of-Line (HOL) blocking, particularly for short decode tasks stuck behind long prefills. \systemname overcomes these limitations by combining dynamic, on-the-fly SM reallocation with out-of-order task batching, effectively mitigating HOL blocking and maximizing SLO attainment.

Aegaeon and AegVoltana benefit from the PD-disaggregated architecture. This design inherently eliminates intra-batch interference, ensuring latency-sensitive decode tasks are not stalled by computation-heavy prefill tasks. In contrast, \systemname utilizes a PD-aggregated (colocated) setup. Consequently, it remains susceptible to minor interference when prefill and decode tasks share the same execution batch, leading to the observed marginal difference in SLO attainment.
\vspace{-4mm}
\subsection{Ablation Study}
\label{s:ablation}
Next, we quantify the contribution of each design part toward energy efficiency, isolating six distinct design features: 

\squishenum{}
\item \textbf{SLO-aware Dispatching:} Route requests based on resource availability and deadline constraints. 
\item \textbf{Frequency-aware Dispatching:} Route requests to match frequency affinities while maintaining SLO compliance. 
\item \textbf{Fine-grained Runtime Management:} Dynamically adjust SM partitioning and GPU frequency. 
\item \textbf{SLO-aware Placement:} Reallocate LLMs across GPUs based on resource demands. 
\item \textbf{Frequency-Affinity Placement (EWR-aware):} Reallocate LLMs based on resource and frequency demands.
\item \textbf{Idle Power Management:} Simulate theoretical energy savings via GPU shutdown or HBM frequency scaling. 
\squishenumend{}

These features are enabled incrementally from Stage-1 to Stage-6. We designate Stage-5 as the standard \systemname configuration, representing the fully implemented system on our testbed. Stage-6 represents \textsf{Festina+}\xspace.
We measure energy consumption under the same workloads in Section~\ref{ss:main_results}.

\begin{figure}[]     
  \centering
  \includegraphics[width=\linewidth]{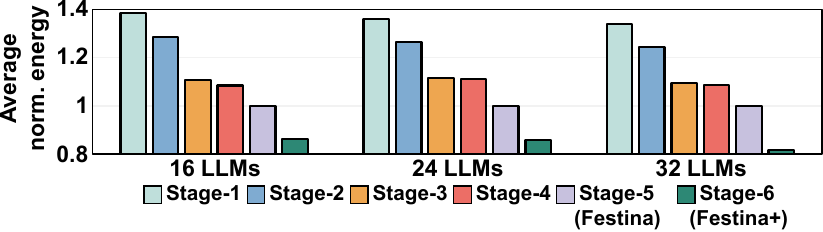}
  \vspace{-8mm}
  \caption{Breakdown of average energy consumption across incremental stages. Results are normalized to Stage-5 (the standard \systemname configuration).}
  \label{fig:ablation}
  \vspace{-5mm}
\end{figure}

Figure~\ref{fig:ablation} details the energy consumption for each stage, normalized to \systemname (Stage-5).
Stage-1 (baseline) exhibits the highest consumption due to the absence of energy-aware optimizations.
Stage-2 introduces frequency-aware dispatching, yielding around 12\% energy reduction. By clustering requests with similar optimal frequencies, this stage prevents the hardware from defaulting to the TDP limit for mismatched workloads.
Stage-3 adds fine-grained runtime management, delivering a substantial 25\% energy saving. 
Stage-4 incorporates standard SLO-aware placement (similar to \cite{dilu,aegaeon}), which offers a modest 2\% energy saving. While this consolidates resources based on demand, it fails to account for phase dominance (i.e., whether energy consumption is dominated by compute-bound prefill or memory-bound decode).
Stage-5 (\systemname) addresses this by enabling EWR-aware placement. By consolidating workloads based on frequency preference, we achieve an additional 11\% energy saving. 
Finally, Stage-6 (\textsf{Festina+}\xspace) demonstrates that if idle GPUs can be fully shut down or HBM frequency can be lowered, a further 16\% energy reduction is achievable.
This points to a promising direction, once such hardware-level controls become feasible.

\subsection{Micro-benchmarks}
\label{ss:sensitivity}
\noindent\textbf{Workload Sensitivity.} We further evaluate \systemname under extreme workload distributions using ShareGPT-ix2 (prefill-heavy) and ShareGPT-ox2 (decode-heavy) from Aegaeon. Figure~\ref{fig:input_length_x2} compares the SLO attainment and energy consumption against the state-of-the-art baseline, Aegaeon.

For the prefill-heavy workload (ShareGPT-ix2), \systemname maintains SLO attainment comparable to Aegaeon while achieving energy savings of 18\% to 30\%. We observe that energy gains diminish as workload intensity (RPS and number of LLMs) increases. This is because high-intensity prefill workloads generate a massive influx of tokens, requiring the GPU to operate near its TDP limit to meet deadlines, thereby reducing the opportunity for frequency downscaling.
Conversely, for the decode-heavy workload (ShareGPT-ox2), \systemname achieves both higher energy efficiency (saving 28\% to 35\% compared to Aegaeon) and better SLO attainment in specific high-load scenarios (e.g., 32 LLMs at 2.5 RPS). 

The performance advantage stems from resource flexibility: Aegaeon relies on static partitioning between prefill and decode nodes, leading to bottlenecks when decode demand outstrips the fixed decode resources. In contrast, \systemname allows decode tasks to opportunistically utilize free SMs from prefill tasks. Regarding energy, Aegaeon relies on the default hardware policy that aggressively boosts frequency to the TDP regardless of phase characteristics. \systemname proactively lowers the GPU frequency for the decode phase, yielding substantial energy savings without violating SLOs.

\begin{figure}[t]     
  \centering
  \vspace{-1mm}
  \includegraphics[width=0.95\linewidth]{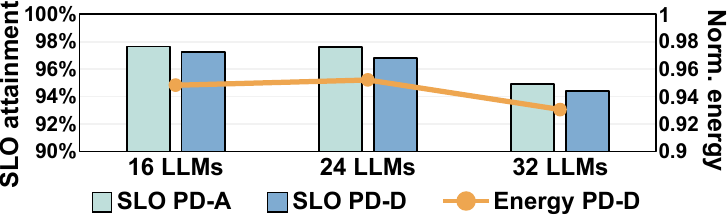}
  \vspace{-4mm}
  \caption{Average SLO attainment and energy consumption (normalized to the PD-aggregated setup) comparing PD-A and PD-D configurations.}
  \label{fig:pd-split}
  \vspace{-8mm}
\end{figure}

\noindent\textbf{Extension to PD-disaggregation.} We further evaluate \systemname in the prefill-decode disaggregated (PD-D) architecture and compare it against the PD-aggregated (PD-A) setup using the same workloads as in \S\ref{ss:main_results}. For PD-D, we adopt Aegaeon’s prefill--decode separation strategy.

Results are presented in Figure~\ref{fig:pd-split}. Regarding SLO attainment, the PD-D setup exhibits a slight degradation of approximately 1.5\% compared to PD-A. This drop is attributed to the overhead of KV cache transmission, which currently lacks low-level optimization in our testbed, delays the initiation of decode tasks, and extends total execution time. However, regarding energy efficiency, the PD-D setup yields an additional 7\% saving over PD-A. This improvement stems from the isolation of workload phases: prefill and decode tasks possess significantly divergent optimal frequencies. By separating them onto dedicated GPUs, the local schedulers can enforce precise frequency tuning for each phase without the interference caused by mixing frequency-mismatched prefill and decode tasks.

\vspace{-2mm}
\section{Related Works}
\label{s:related_works}

\noindent\textbf{High-Performance \& Spatial Sharing Systems.} Prior systems like ~\cite{10.1145/3624062.3624238,10254957} show optimal GPU sharing while ~\cite{wangMIGRator,dilu,aegaeon,wangwages} optimize latency and GPU utilization. ServerlessLLM~\cite{fu2024serverlessllm} and Medusa~\cite{serverlessllm-medusa} improve auto-scaling by reducing cold-start and checkpoint-migration latency. MuxServe~\cite{muxserve}, Prism~\cite{yu2025prism}, Dilu~\cite{dilu}, and Aegaeon~\cite{aegaeon} enable \textit{spatial sharing} through model co-location and dynamic resource allocation, ranging from whole GPUs to fine-grained SM partitions. These systems prioritize throughput, SLO attainment, and utilization, without explicitly targeting energy efficiency or using frequency scaling to reduce power.

\noindent\textbf{Energy-aware serving systems.}
Recent systems such as DynamoLLM~\cite{dynamollm}, GreenLLM~\cite{liu2025greenllm}, and VoltanaLLM~\cite{yu2025voltanallm} use DVFS to reduce LLM serving energy. However, they are designed around assumptions that do not hold in serverless multi-tenant serving. First, they largely assume that a GPU is dedicated to one model or one serving pipeline. Energy optimization can therefore be reduced to choosing a frequency based on workload intensity, such as input/output length or serving phase. In contrast, \systemname targets a serverless regime where multiple tenants may share one GPU. Co-location couples the energy and latency behavior of different requests and makes frequency selection a shared scheduling decision rather than a per-model control knob.

Second, prior DVFS systems do not deal with shared-frequency contention of co-resident models.
GPU clock may be pinned by the most demanding tenant, forcing other requests to run at a higher-than-needed frequency and wasting energy. 
\systemname reduces the waste from frequency mismatch through frequency-aware dispatching, which routes requests to GPUs with clocks closest to their optimal frequencies.


Third, co-location expands the control space beyond single-model DVFS: frequency selection must be coordinated with SM partitioning across tenants. The joint configuration space grows rapidly with the number of tasks, making exhaustive per-batch search impractical. \systemname uses a lightweight frequency sweep and phase-aware $(\text{SM}, \text{frequency})$ adaptation to keep scheduling overhead low while accounting for distinct energy-optimal operating points of prefill and decode.

These differences are structural rather than missing features that can be added incrementally. DynamoLLM relies on fixed per-pool frequencies from offline profiling and rigid per-LLM pools, which conflict with dynamic serverless co-location. GreenLLM and VoltanaLLM introduce phase-specific frequency control, but they rely on PD-disaggregated serving with static prefill/decode partitions and do not provide auto-scaling or multi-tenant spatial multiplexing. Adapting these systems to serverless serving would require co-location-aware dispatch, shared-frequency conflict resolution, joint SM/frequency adaptation, and frequency-aware scale-in or migration. These mechanisms are central to \systemname rather than extensions to prior designs. Table~\ref{tab:comparison} summarizes the comparison between \systemname and prior systems.

\vspace{-3mm}
\section{Conclusion}
We present the design, implementation, and evaluation of \systemname, a power-efficient serverless LLM serving system. \systemname leverages a global scheduler and per-GPU local scheduler to jointly optimize the request placement and GPU resources to minimize the energy consumption of LLM serving without violating each request's latency SLOs.
Moreover, \systemname leverages an EWR-aware mechanism that consolidates workloads via SLO-aware migration to further reduce static memory power consumption.
Evaluations show that \systemname can save up to 56\% energy compared to existing SOTA LLM serving systems while maintaining SLO attainment within a 2\% margin.

\vspace{-2mm}
\section*{Acknowledgement}
We appreciate the feedback from our shepherd and the anonymous reviewers. 
This work is supported by the National Science Foundation under Grant No.2337537 and No.2302724.

\bibliographystyle{ACM-Reference-Format}
\bibliography{reference}
\newpage
\appendix

\section{Background}
\label{s:background}

\subsection{LLM Inference}
Contemporary LLMs~\cite{llama_1, gemini, chatgpt} are predominantly decoder-only Transformers composed of repeated blocks that consist of multi-head attention and feed-forward layers.
Given input prompts, LLMs generate tokens in an autoregressive manner.
The LLM inference splits into two distinct phases: prefill and decode.

In the prefill phase, the input prompt is processed once to build the key-value (KV) cache~\cite{kwon2023efficient}, an optimization to eliminate redundant computation of key-value matrix contents across decoding iterations, and emit the first output token. The prefill phase, characterized by long input prompts, heavily utilizes computation resources (i.e., SMs). At the end of the prefill phase, a new token is generated, then used as the input in the decode phase.
In the decode phase, the LLM generates one new token per iteration until it reaches a predefined sequence length or the end-of-sentence marker. Each iteration reuses the stored KV cache. This KV cache optimization typically shifts the bottleneck from computation to memory bandwidth, making the decode phase memory-bound. 
Moreover, Time-to-First-Token (TTFT) captures latency from request arrival to the first output token, used as the performance metric for the prefill phase. Time-Between-Tokens (TBT) captures latency between successive output tokens, used as the performance metric for the decode phase.

\subsection{Serverless LLM Serving}
Conventional server-centric LLM deployments have large resource footprints and require extensive server management. In contrast, cloud vendors and researchers have advanced serverless serving systems, which elastically scale inference workers to match demand. With pay-as-you-go pricing—charging only for the time each worker runs—serverless is cost-effective for bursty, long-tail workloads~\cite{serverless-asyfunc,serverless-optimus,serverless-infless}.

Recent serverless LLM solutions~\cite{serverlessllm-medusa,serverlessllm-towards, fu2024serverlessllm} focus on reducing cold-start latency, incurred when no worker hosts the target LLM and a new one must be initialized. However, they still rely on static, exclusive GPU allocations optimized for the most demanding workloads, which leads to underutilization, GPU waste, and energy inefficiency for lighter workloads with shorter inputs.

\subsection{GPU Sharing}
Modern GPUs support two advanced sharing mechanisms: (i) Multi-Process Service (MPS)~\cite{nvidia_mps} and (ii) Multi-Instance GPU (MIG)~\cite{nvidia-mig-docu}.
MPS partitions compute resources (SMs) to reduce compute contention, while keeping memory shared, making it suitable for memory-hungry workloads like LLMs. It also supports runtime SM repartition without disrupting running models.
MIG partitions both compute and memory at the GPC level, providing full physical isolation. However, it offers only 5 fixed partition sizes (1, 2, 3, 4, or 7 GPCs), often with too little memory for LLMs, and reconfiguration is possible only when involved GPCs are idle, incurring checkpointing and initialization overheads~\cite{spatialmig-miso,zhang2023migperf,li2024star,wangMIGRator}.
Due to MIG’s coarse granularity, limited memory, and costly reconfiguration, MPS has emerged as the preferred mechanism for GPU-shared LLM serving~\cite{muxserve,dilu,aegaeon}. We align with this established paradigm and employ MPS in this work.

\subsection{Dynamic Voltage and Frequency Scaling}
Dynamic voltage and frequency scaling (DVFS) is a prevalent energy management technique for CPUs and GPUs to balance performance and power consumption.
By adjusting operating voltage and frequency, DVFS exploits the physical relationship where power consumption ($P$) is proportional to the square of the voltage ($V$) and the frequency ($f$), expressed as $P \propto V^2f$~\cite{power_voltage_frequency,huawei-dvfs}.
While CPUs often benefit from mature DVFS governors, GPU-based strategies remain comparatively under-optimized~\cite{huawei-dvfs,dvfs-joss,dvfs-geepafs}. 
Furthermore, applying DVFS in the context of LLM serving poses a unique challenge: energy efficiency strategies must be implemented without compromising SLOs.

\begin{figure}[t]
    \centering  
    \includegraphics[width=1\columnwidth]{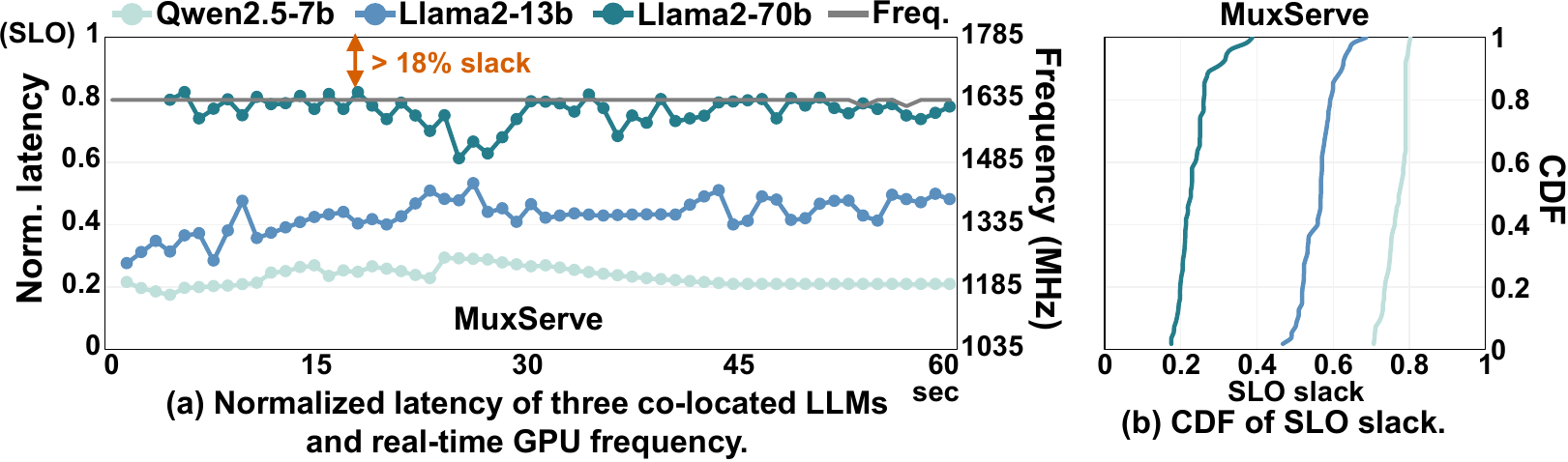}
    \vspace{-6mm}
    \caption{(a) Normalized latency of three co-located LLMs and real-time GPU frequency. (b) CDF of SLO slack for three LLMs during serving requests.}
    \label{fig:motivation_llama2_70b}
    \vspace{-6mm}
\end{figure}

\begin{figure*}[t]     
  \centering
  \includegraphics[width=\linewidth]{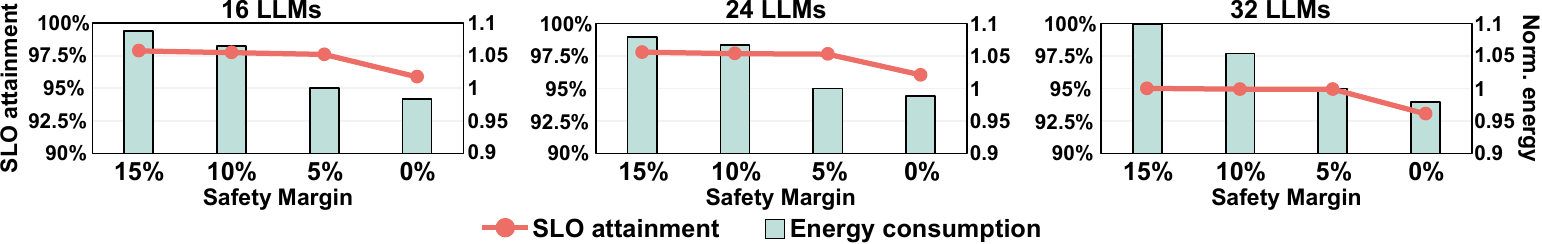}
  \caption{Impact of varying safety margins on average SLO attainment and energy consumption. Energy values are normalized to the 5\% margin baseline. Same Workloads in Section~\ref{ss:main_results} are utilized.}
  \label{fig:safety_margin}
\end{figure*}

\section{SLO Slack for Large LLMs}
\label{ss:slack_large_llms}

We utilize the same trace described in Section~\ref{sec:motivation_subsection_energy_opportunity} and employ MuxServe as the serving platform with two H100s due to memory space shortage. As illustrated in Figure~\ref{fig:motivation_llama2_70b}, even when co-locating large LLMs such as Llama2-70b, we observe a minimum of 18\% SLO slack. This confirms that the opportunity to enhance energy efficiency by lowering frequency remains viable, even when serving substantial models. We observe consistent results when utilizing Aegaeon as the serving platform.

\section{Energy Waste from Frequency Mismatch}
\label{ss:energy_waste_freq_mismatch}
Co-locating multiple LLMs improves utilization and enables scale-in, but different models exhibit different energy-optimal frequencies. Because GPU frequency is effectively device-wide and often pinned by the most demanding workload, frequency-agnostic packing can force low-frequency workloads to run at an unnecessarily high clock, wasting energy (e.g., we observe up to 10--16\% overhead under mismatched co-location for the same models in \S\ref{sec:necessity_of_reconfiguration}). Therefore, cluster-level placement and consolidation must explicitly account for frequency affinity.

\section{Scheduling Latency Breakdown}
\label{ss:scheduling_overhead}

We evaluate the scalability of \systemname's control plane by measuring the scheduling latency of both the global and local schedulers across varying cluster sizes and queue depths (i.e., number of pending tasks). 
Our simulations, which isolate algorithmic overhead by excluding network variance, demonstrate that \systemname remains highly responsive at scale. 
As detailed in Table~\ref{tab:scheduling_latency}, the global scheduler maintains a dispatch latency of under 1 ms even when managing a cluster of 10,000 GPUs. 
Similarly, the local scheduler determines the next task batch in less than 0.1 ms, even with 100 tasks in the queue. 
These results confirm \systemname’s robustness and ability to handle massive-scale deployments with negligible overhead.

\begin{table}[h]
    \centering
    \caption{Scheduling latency of \systemname's control plane (global scheduler and local scheduler) under various cluster sizes and queue sizes.}
    \label{tab:scheduling_latency}
    \vspace{0.2cm}
    \small
    \begin{tabular}{cccc}
        \toprule
        \multicolumn{2}{c}{\textbf{Global Scheduler}} & \multicolumn{2}{c}{\textbf{Local Scheduler}} \\
        \multicolumn{2}{c}{\textit{(Linear Scan)}} & \multicolumn{2}{c}{\textit{(Priority Batching)}} \\
        \cmidrule(lr){1-2} \cmidrule(lr){3-4}
        \textbf{Cluster Size} & \textbf{Latency} & \textbf{Queue Size} & \textbf{Latency} \\
        \textbf{(GPUs)} & \textbf{(ms)} & \textbf{(Tasks)} & \textbf{(ms)} \\
        \midrule
        100 & 0.011 & 10 & 0.0544 \\
        1,000 & 0.103 & 25 & 0.0612 \\
        5,000 & 0.509 & 50 & 0.0768 \\
        10,000 & 0.988 & 100 & 0.0942 \\
        \bottomrule
    \end{tabular}
\end{table}

\section{Efficient State Transfer Across Different GPUs}
\label{ss:design_transfer}

After the global controller computes a new placement, the system needs to (re)initialize the inference engine and move the necessary execution state (model weights and, for in-flight requests, prompt/KV state) before normal serving can resume. Because these states can be large, naïvely transferring everything can dominate reconfiguration time and visibly disrupt inference.

\noindent\textbf{Transmission plan (device mapping).}
Motivated by prior work on minimizing reconfiguration communication cost (e.g., SpotServe \cite{miao2024spotserve}), we compute an explicit mapping from \emph{source} GPUs (current placement) to \emph{destination} GPUs (target placement) that minimizes total transfer time. We model the mapping as a weighted bipartite assignment problem: each edge $(u,v)$ represents transferring the state currently on GPU $u$ to GPU $v$, with cost
\[
e_{uv} = \frac{\text{Bytes}(u \rightarrow v)}{\text{BW}(u,v)},
\]
where $\text{Bytes}(\cdot)$ accounts for the model weights and any request state selected for migration, and $\text{BW}(u,v)$ is the measured interconnect bandwidth. We then solve the minimum-cost assignment using the Kuhn--Munkres (Hungarian) algorithm~\cite{kuhn1955hungarian} to obtain a migration plan with low aggregate transfer overhead. 

\noindent\textbf{Selective request migration (finish locally vs.\ move).}
To avoid unnecessary transfers, we adopt a simple cost-based rule, similar in spirit to prior live-migration systems for LLM inference: for each in-flight request, we compare the estimated \emph{remaining computation time} if it finishes on the source GPU against the estimated \emph{migration time} under the chosen mapping~\cite{mell, fu2024serverlessllm}.
Requests that are close to completion are allowed to finish in place, while requests with substantial remaining work are migrated, reducing both network traffic and reconfiguration disruption.

\noindent\textbf{What to transfer (tokens/prompt vs.\ KV cache).}
For requests selected for migration, we further choose between two state-transfer modes. Following the same principle as ServerlessLLM and related work, we may transfer only the minimal token state (input prompt and generated tokens) and recompute the KV cache on the destination, or transfer both tokens and KV cache when the link is fast and recomputation would be slower~\cite{miao2024spotserve}.
This decision is made per request using lightweight time estimates for (i) KV recomputation and (ii) KV transmission, favoring the cheaper option.

\noindent\textbf{Staged, memory-safe transfer.}
Finally, to prevent out-of-memory (OOM) on the destination GPU during reconfiguration, we perform transfers in stages using a simple reservation/credit protocol inspired by the coordinated, multi-stage migration style in Llumnix~\cite{Llumnix}.
The destination scheduler first reserves memory for locally retained requests, then admits incoming transfers only when sufficient free space is available. As local execution frees memory, the destination releases additional credits, allowing the source to continue sending state incrementally. This staged approach bounds transient memory pressure and avoids service stalls caused by OOM during migration.

\section{Impact of Safety Margin}
\label{ss:impact_of_safety_margin}

The safety margin serves as a resource buffer during placement reconfiguration, designed to absorb minor workload fluctuations and prevent rapid oscillation between scale-in and scale-out operations. Figure~\ref{fig:safety_margin} illustrates the trade-off between SLO attainment and energy consumption across margins ranging from 0\% to 15\%. Regarding SLO attainment, a 0\% margin is insufficient, resulting in a 2\% performance drop compared to a 5\% margin. In contrast, margins of 5\%, 10\%, and 15\% yield comparable reliability, with performance variances below 0.5\%. 

Regarding energy, higher margins (10\% and 15\%) result in higher consumption, which strict resource reservation restricts opportunities for workload consolidation. While the 0\% margin achieves the lowest energy consumption (approximately 2\% lower than the 5\% setting), at the cost of SLO violations. Consequently, we empirically select a 5\% safety margin as the optimal equilibrium between SLO adherence and energy efficiency.

\section{Energy Consumption during Migration}
\label{ss:migration_energy_consumption}
We measured migration energy on our H100/NVLink testbed. Migrating Llama2-13B with a 4K-token KV cache costs 10.67 J, recovered after 108 ms of avoided idle power; migrating Qwen3-32B costs 40.76 J, recovered in 413 ms. Over a 5-minute consolidation window, disabling one GPU saves 29.6 kJ: 2788$\times$/727$\times$ the two migration costs.

\section{Prior Energy-Efficient LLM Serving Systems}
\label{ss:related_energy_efficieint_works}
Recent research has increasingly focused on improving the energy efficiency of LLM serving.
We explain these representative systems and highlight their difference with \systemname.

\noindent\textbf{DynamoLLM~\cite{dynamollm}.} This system targets server-centric, single-LLM scenarios. It classifies requests into groups based on input and estimated output lengths. It then assigns a static, energy-optimal frequency to each group to minimize energy consumption.

\noindent\textbf{GreenLLM~\cite{liu2025greenllm}.} GreenLLM is particularly designed for a PD-disaggregated setup. It leverages offline profiling to dynamically adjust frequencies for prefill batches and employs a feedback-driven mechanism to adjust decode frequencies to save energy.

\noindent\textbf{VoltanaLLM~\cite{yu2025voltanallm}.} Similar to GreenLLM, VoltanaLLM targets single-LLM serving under PD-disaggregated setups. It independently adjusts the frequency for prefill and decode nodes based on current workloads, aiming to minimize energy usage while satisfying SLO constraints.

\noindent\textbf{Limitations in Serverless Contexts.} While these works effectively address energy efficiency in dedicated environments, they are not well-suited for shared, serverless scenarios due to five critical limitations:

\textbf{1. Lack of Auto-scaling:} GreenLLM and VoltanaLLM rely on static node partitioning for prefill and decode phases in disaggregated setups. They lack the dynamic auto-scaling capabilities required for serverless serving. Without scale-out support, meeting SLOs under high load is difficult; conversely, without scale-in support, GPUs remain under-utilized during low load, wasting resources and preventing other tenants from occupying GPUs.

\textbf{2. Absence of Spatial Multiplexing:} These systems do not support spatial multiplexing for co-located multi-tenancy. Dedicating a single high-performance GPU (e.g., NVIDIA H100) to one LLM often results in severe under-utilization~\cite{muxserve,aegaeon,dilu}. However, na\"ive sharing without dedicated resource management (e.g., SM partitioning, frequency control) leads to significant interference and SLO violations~\cite{dhakal2020gslice,laius,gpu-let}. The lack of fine-grained management limits the effectiveness of these systems in shared environments.

\textbf{3. No Shared-Frequency Conflict Resolution:} Prior systems assume each frequency setting governs a single workload. Under co-location, all tenants on a GPU share one device-wide clock, so the most demanding co-tenant pins the frequency and forces models that prefer lower frequencies to over-clock, eroding DVFS savings. None of these systems provide a dispatching mechanism that groups frequency-compatible tenants to avoid this conflict.

\textbf{4. Inflexible Phase Management:} DynamoLLM applies a single static frequency for both prefill and decode phases, neglecting their distinct computational intensities. While GreenLLM and VoltanaLLM provide phase-specific adjustments, they are tailored for disaggregated setups and do not support prefill-decode mixed scenarios, where PD-aggregated serving is common in current serverless scenarios~\cite{serverlessllm-medusa,fu2024serverlessllm,dilu,serverless-enova,serverless-flexpipe}.

\textbf{5. Decoupled, Not Joint, Control:} These systems treat frequency as the sole energy knob. In a shared GPU, SM partitioning and frequency are coupled---the SM allocation shifts each tenant's latency/energy-optimal frequency---so optimizing either in isolation is sub-optimal. Prior controllers offer no mechanism for the joint SM/frequency adaptation that co-located serving demands.


\end{document}